\documentclass[11pt,twocolumn,tight,times]{aastex62}
\usepackage{graphicx,color}
\usepackage{subfigure}
\usepackage{mathrsfs,amsmath}
\usepackage{natbib}
\usepackage{ulem}
\usepackage{hyperref}
\usepackage{CJK}
\emergencystretch=\maxdimen
\begin{document}
\begin{CJK*}{UTF8}{gbsn}

\title{Constraining the Binarity of Massive Black Holes in the Galactic Center and Some Nearby Galaxies via Pulsar Timing Array Observations of Gravitational Waves}
\shorttitle{Constraining Binarity of Nearby MBHs by PTA Observations}
\shortauthors{Guo, Yu, \& Lu}
\author[0000-0001-5174-0760]{Xiao Guo}
\affil{
National Astronomical Observatories, Chinese Academy of Sciences, 20A Datun Road, Beijing 100101, China}
\affil{
School of Astronomy and Space Science, University of Chinese Academy of Sciences, 19A Yuquan Road, Beijing 100049, China
}
\affiliation{School of Fundamental Physics and Mathematical Sciences, Hangzhou Institute for Advanced Study, University of Chinese Academy of Sciences, Hangzhou 310024, China}
\author[0000-0002-1745-8064]{Qingjuan Yu$^{*}$}
\affil{
Kavli Institute for Astronomy and Astrophysics, and School of Physics, Peking University, No. 5 Yiheyuan Road, Beijing 100871, China
}
\author[0000-0002-1310-4664]{Youjun Lu$^{\dagger}$}
\affil{
National Astronomical Observatories, Chinese Academy of Sciences, 20A Datun Road, Beijing 100101, China}
\affil{
School of Astronomy and Space Science, University of Chinese Academy of Sciences, 19A Yuquan Road, Beijing 100049, China
}
\email{$^*$\,yuqj@pku.edu.cn}
\email{$^\dagger$\,luyj@nao.cas.cn}


\begin{abstract}
Massive black holes (MBHs) exist in the Galactic center (GC) and other nearby galactic nuclei. As natural outcome of galaxy mergers, some MBHs may have a black hole (BH) companion. In this paper, assuming that the MBHs in the GC and some nearby galaxies are in binaries with orbital periods ranging from months to years (gravitational-wave frequency $\sim1-100$\,nHz), we investigate the detectability of gravitational-waves from these binary MBHs (BBHs) and constraints on the parameter space for the existence of BBHs in the GC, LMC, M31, M32, and M87, that may be obtained by current/future pulsar timing array (PTA) observations. We find that a BBH in the GC, if any, can be revealed by the Square Kilometer Array PTA (SKA-PTA) if its mass ratio $q\gtrsim10^{-4}-10^{-3}$ and semimajor axis $a\sim20-10^3$\,AU. The existence of a BH companion of the MBH can be revealed by SKA-PTA with $\sim20$-year observations in M31 if $q\gtrsim10^{-4}$ and $a\sim10^2-10^4$\,AU or in M87 if $q\gtrsim10^{-5}$ and $a\sim10^3-2\times10^4$\,AU, but not in LMC and M32 if $q\ll1$. If a number of milli-second stable pulsars with distances $\lesssim0.1-1$\,pc away from the central MBH in the GC, LMC, M32, or M31, can be detected in future and applied to PTAs, the BH companion with mass even down to $\sim100M_\odot$, close to stellar masses, can be revealed by such PTAs. Future PTAs are expected to provide an independent way to reveal BBHs and low-mass MBH companions in the GC and nearby galaxies, improving our understandings of the formation and evolution of MBHs and galaxies.
\end{abstract}


\section{Introduction}
\label{sec:intro}

The Galactic center (GC), hosting a massive black hole (MBH)  \citep{1974ApJ...194..265B, 2002Natur.419..694S, 2003ApJ...586L.127G, 2008ApJ...689.1044G, RevModPhys.82.3121}, is a unique laboratory to test the general relativity and study the interaction of the MBH with its surrounding environment and formation of galactic nuclei \citep[e.g.,][]{ALEXANDER200565, 2020Innov...100063Y}. It was conjectured that an intermediate-mass black hole (IMBH) may exist in the GC and rotate around the GC MBH to form a low-mass-ratio massive binary black hole (BBH; see \citealt{2003ApJ...599.1129Y, 2003ApJ...593L..77H, 2006ApJ...641..319P, 2007ApJ...666..919Y, RevModPhys.82.3121, 2017ApJ...850L...5T, 2019MNRAS.482.3669G, 2019ApJ...871L...1T}). A number of studies investigated the effects of such a hypothetical IMBH on the motion of Sgr A* or S2/S0-2 and found that its existence may be allowed within some parameter space (\citealt[][]{2003ApJ...599.1129Y, 2003ApJ...593L..77H, RevModPhys.82.3121, 2019MNRAS.482.3669G, 2020ApJ...888L...8N, 2020A&A...636L...5G, 2023A&A...672A..63G, 2023ApJ...959...58W}). Nevertheless, if such an IMBH does exist with an orbital period in the range from months to years, its rotation around the GC MBH can lead to gravitational wave (GW) radiation at frequency nHz-$\mu$Hz, potentially serving as a probe to the binary orbital and physical properties.

The GW signal (nHz-$\mu$Hz) emitted from individual BBHs in galactic centers \citep[e.g.,][]{1980Natur.287..307B, 2002MNRAS.331..935Y} may be extracted from the timing series data of multiple pulsars monitored by Pulsar Timing Arrays (PTAs) \citep[e.g.,][]{1978SvA....22...36S,1979ApJ...234.1100D,maggiore2008gravitational, 2009MNRAS.394.2255S, 2010CQGra..27h4016S, 2011gwpa.book.....C, 2011MNRAS.414.3251L, manchester2013pulsar, 2013CQGra..30x4009S, 2014gwdd.book.....V, doi:10.1093/mnras/stu1717, 2015SCPMA..58.5748B,  mingarelli2015gravitational, 2016MNRAS.459.1737S, 2019BAAS...51c.336T,2021arXiv210513270T,2023ApJ...951L..50A, 2023arXiv230616226A}. Currently a number of PTA experiments are operating, including the Parkes PTA (PPTA\footnote{\url{http://www.atnf.csiro.au/research/pulsar/ppta/}}; \citealt{manchester2013pulsar}), the European PTA (EPTA\footnote{\url{http://www.epta.eu.org/}}; \citealt{2013CQGra..30v4009K}), the North American Nanohertz Observatory for Gravitational Waves (NANOGrav\footnote{\url{http://nanograv.org/}}; \citealt{2013CQGra..30v4008M, 2019BAAS...51g.195R}), the Indian Pulsar Timing Array (InPTA; \citealt{2018JApA...39...51J}),  the Chinese PTA (CPTA; \citealt{2011IJMPD..20..989N, 2009A&A...505..919S}), and the MeerTime PTA (MPTA; \citealt{2023MNRAS.519.3976M}). The former three have collected data for more than a decade to search for possible low frequency GW signal and were also combined together to form the International PTA by sharing the data (IPTA\footnote{\url{http://www.ipta4gw.org/}}; \citealt{2013CQGra..30v4010M,10.1093/mnras/stw347, 2019MNRAS.490.4666P}). In the future, the Square Kilometer Array (SKA) \citep[e.g.,][]{Lazio2013SKA,  2017PhRvL.118o1104W} is expected to find some new stable millisecond pulsars (MSPs) and establish a PTA with high sensitivity, denoted as SKA-PTA hereafter, to detect low frequency GW sources. 

Recently, CPTA, EPTA(+InPTA), NANOGrav, and PPTA reported the evidence for the existence of the nHz gravitational wave background (GWB) with a confidence level of $\sim2-4\sigma$ \citep[][see also \citealt{2020ApJ...905L..34A}]{2023RAA....23g5024X, 2023ApJ...951L...8A, 2023ApJ...951L...6R, 2023A&A...678A..50E}. This GWB can be interpreted as the GWs from the cosmic inspiralling BBHs, suggesting that the detection of individual BBHs could soon be realized \citep[e.g.,][]{2023ApJ...952L..37A, 2023arXiv230616227A, 2023ApJ...955..132C, CYL2024} and the binarity of MBHs in the centers of some nearby galaxies may be also revealed by PTAs.

In this paper, we investigate whether the GW signal radiated from a hypothetical IMBH rotating around the GC MBH can be detected by PTAs and how strongly a constraint can be put on the mass and the orbital parameter space of such an IMBH by future PTAs with higher sensitivities. We also further investigate the possibility of revealing the binarity of MBHs in the centers of several nearby typical galaxies, i.e., large Magellanic cloud (LMC; dwarf irregular satellite), M31 (disk galaxy), M32 (dwarf elliptical galaxy), and M87 (giant elliptical galaxy), and their corresponding BBH parameter space via PTAs. Exploration of the MBH binarity and its mass ratio, and the lower-mass-limit of the companion in the various types of galaxies is one of the important parts to expand our horizon of GW source detections and our understandings of the formation and evolution of MBHs and galactic nuclei.

This paper is organized as follows. In Sections~\ref{sec:GWsignal} and \ref{sec:GWstrain}, we introduce the basic formulas to estimate the GW signal from a BBH and its characteristic strain, respectively. In Section~\ref{sec:PTA}, we introduce the framework to estimate the expected signal-to-noise ratio (S/N) of an eccentric BBH monitored by a PTA. In Section~\ref{sec:GCMBBH}, we estimate the expected S/N of a hypothetical BBH in the GC with various masses, mass ratios, and eccentricities, and investigate the parameter space of the MBH companion or IMBH whose existence can be revealed independently by PTA observations. In Section~\ref{sec:otherMBBH}, we investigate whether future PTAs can reveal or put constraints on the binarity of those MBHs in a few nearby typical galaxies, i.e., LMC, M31, M32, and M87. Our main conclusions are summarized in Section~\ref{sec:conclusion}.

\section{GW signals from a hypothetical BBH in the GC}
\label{sec:GWsignal}

Assume that a massive BBH is located with mass $M_{\rm BBH}$ at the GC is a BBH, where we adopt the value of   $M_{\rm BBH}$ as $M_{\rm BBH} = 4.26\times 10^6 M_\odot$  \citep[][see also \citealt{2002Natur.419..694S,2003ApJ...586L.127G, 2003ApJ...596.1015S,  2008ApJ...689.1044G,  2016ApJ...830...17B, 2017ApJ...837...30G, 2019Sci...365..664D}]{2020A&A...636L...5G}. This BBH is assumed to be on an orbit with eccentricity $e$ and semimajor axis $a$ (much larger than the gravitational radius $GM_{\rm BH,1}/c^2$), where $G$ is the gravitational constant, $c$ is the speed of light, and $M_{\rm BH,1}$ is the mass of the primary component. The mass ratio of the secondary component ($M_{\rm BH,2}=M_{\rm BBH}-M_{\rm BH,1}$) to the primary one is $q = M_{\rm BH,2} / M_{\rm BH,1} \ll 1$. The distance of this BBH to the Earth is $8$\,kpc \citep{RevModPhys.82.3121, ALEXANDER200565,2019Sci...365..664D}. By assuming that the inclination angle between the normal of the BBH orbital plane and the line of sight is $\iota=0\arcdeg$,\footnote{The general dependence of the received GW signals on $\iota$ is included in the geometric factor $\chi$, see Equations~\eqref{eq:SNR_MFc}-\eqref{eq:SNR_MFe}, \eqref{eq:SNR_CCc}-\eqref{eq:SNR_CCe} below. See detailed formulas of $\chi$ in \citet[][]{2022ApJ...939...55G}.} the GW strain in the time domain radiating from this hypothetical BBH is given by \citep{Jaranowski:2009zz, 2011gwpa.book.....C, maggiore2008gravitational}
\begin{eqnarray}
h_a &=&\frac{G^2M_{\rm BBH}\mu}{c^4 a r}\Psi_a(\phi,e) \nonumber  \\
& = & 1.08\times10^{-14} \left(\frac{M_{\rm BBH}}{4.26\times 10^6M_\odot}\right)^2 \left(\frac{100{\rm AU}}{a}\right) \left( \frac {8{\rm kpc}}{r}\right) \nonumber \\
& & \times f(q)\Psi_a(\phi,e).  
\label{eq:ha}
\end{eqnarray}
where $a=+,\times$, and
\begin{equation}
\Psi_+(\phi,e)=-\frac{4\cos2\phi+5e\cos\phi+e\cos3\phi+2e^2}{1-e^2},
\label{eq:psiplus}
\end{equation}
\begin{equation}
\Psi_\times(\phi,e)=-\frac{4\sin2\phi+5e\sin\phi+e\sin3\phi}{1-e^2}.
\label{eq:psiminus}
\end{equation}
Here $h_+$ and $h_\times$ are the plus and the cross polarizations of the GW, $\mu=M_{\rm BH,1}M_{\rm BH,2}/M_{\rm BBH}=\frac{q}{(1+q)^2}M_{\rm BBH}$ is the reduced mass, $f(q) = \frac{q}{(1+q)^2}$ ($\sim q$ if $q\ll1$), and $\phi$ is the BBH orbital phase angle. In general, the change of $\phi$ with time $t$ should be obtained by solving those equations that govern the orbital decay of eccentric binaries under the GW radiation. However, for all the BBH systems considered in this paper, the timescale for orbital decay caused by the GW radiation is much longer than the orbital period, and thus for simplicity and for the purpose of this paper we adopt the Newtonian approximation for the orbital motion and obtain the change of $\phi$ with time by solving
\begin{equation}
\frac{d\phi}{dt} = \frac{\sqrt{GM_{\rm BBH}} (1+e\cos \phi)^2}{a^{3/2}(1-e^2)^{3/2}}.
\end{equation}

\begin{figure}
\centering
%
\includegraphics[width=0.48\textwidth]{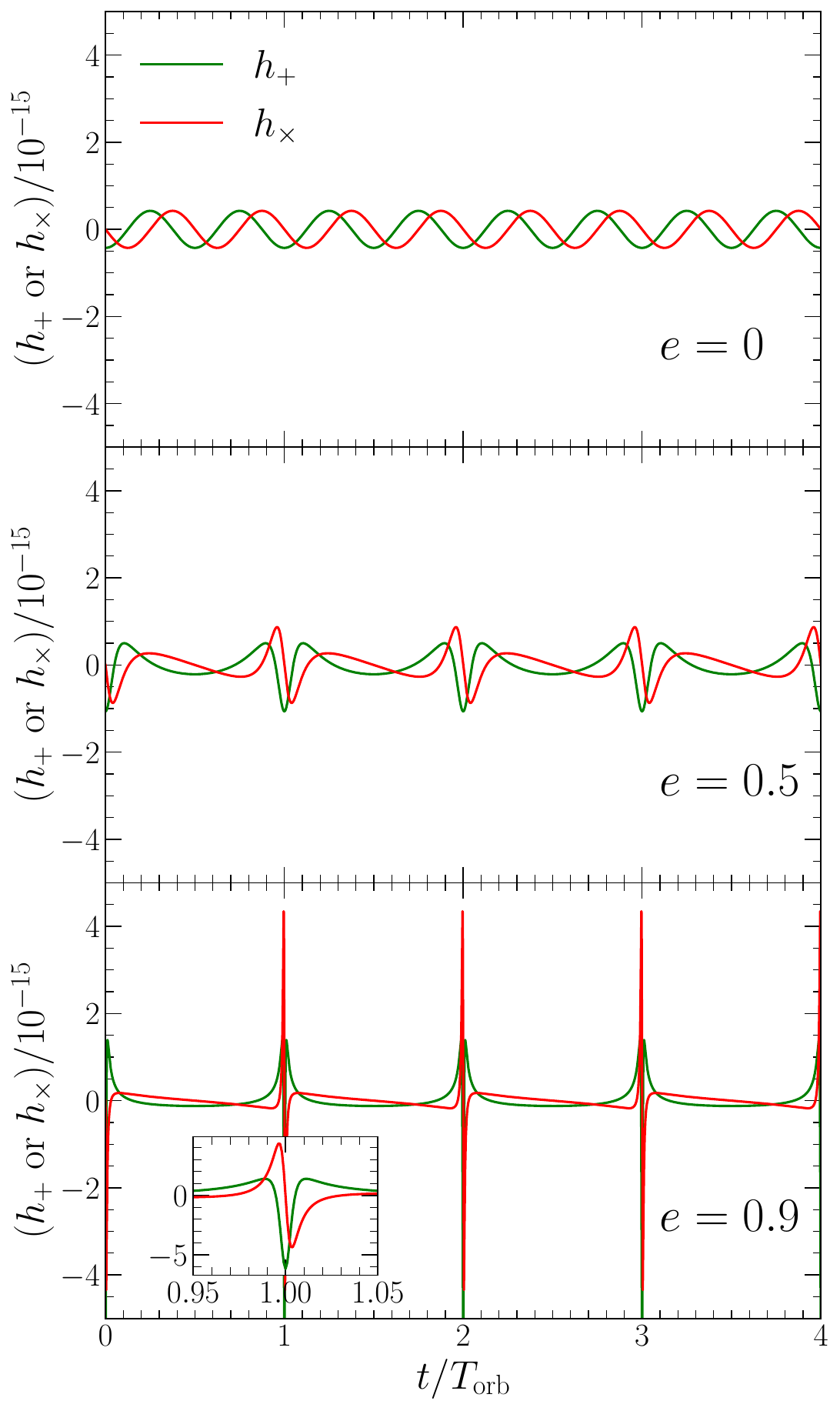}
\caption{The GW waveform radiated from a hypothetical BBH on a Keplerian orbit in the GC with total mass $M_{\rm BBH} = 4.26\times 10^6 M_\odot$, mass ratio $q=0.01$, $\iota=0^{\rm o}$, semimajor axis $a=100$\,AU, eccentricity $e=0$ (top panel), $e=0.5$ (middle panel), or $e=0.9$ (bottom panel) (see Eqs.~\ref{eq:ha}-\ref{eq:psiminus}).
}
\label{fig:f1}
\end{figure}

Figure~\ref{fig:f1} shows some examples for the GW waveform emitted from a hypothetical BBH in the GC, with $q=0.01$, $a=100{\rm AU}$, $e=0$ (top panel), $0.5$ (middle panel), or $0.9$ (bottom panel). As shown in this figure, with increasing eccentricity $e$, the amplitude of the GW strain emitted at the pericenter increases, but the time duration for the peak GW radiation amplitude becomes shorter. This suggests that the GW signal from a highly eccentric BBH may not be easier to be detected compared with that from a circular BBH with the same other physical parameters as naively expected, because a significantly higher cadence is required for the former one. For BBHs with period longer than a few years, however, it may be possible to detect them if they are highly eccentric as the  higher order harmonic GW emission enter into the detection band of PTAs.

In the frequency domain, the angle ($\iota$)-averaged GW power from an eccentric BBH can be expressed as 
\begin{equation}
P=\sum_{n=1}^{\infty} P(n),
\end{equation}
where $P(n)$ ($n=1, 2, 3, \cdots$) is the GW power radiated in the $n$th harmonic from an elliptic BBH given by \citep{2015PhRvD..92f3010H, maggiore2008gravitational}
\begin{equation}
P(n)=\frac{32G^4 M_{\rm BH,1}^2M_{\rm BH,2}^2 M_{\rm BBH}}{5 c^5 a^5}g(n,e),
\end{equation}
\begin{equation}
\begin{split}
g(n,e)=&\frac{n^4}{32}\bigg[\Big\{J_{n-2}(ne)-2eJ_{n-1}(ne)\\
&+\frac{2}{n}J_n(ne)+2eJ_{n+1}(ne)-J_{n+2}(ne)\Big\}^2\\
&+(1-e^2)\Big\{J_{n-2}(ne)-2J_n(ne)+J_{n+2}(ne)\Big\}^2\\
&+\frac{4}{3n^2}J^2_n(ne)\bigg],
\end{split}
\end{equation}
and $J_n$ is the $n$th order Bessel function. The root-mean-square (RMS) amplitude of the angle-averaged GW strain in the $n$th harmonic is \citep{2015PhRvD..92f3010H}
\begin{equation}
H_{n,\rm ave}=2\sqrt{\frac{32}{5}g(n,e)}\frac{G^{5/3}\mathcal{M}^{5/3}}{c^4r n}(2\pi f_{\rm orb})^{2/3},
\end{equation}
where the chirp mass $\mathcal{M}=\nu^{3/5}M_{\rm BBH}$, $\nu=M_{\rm BH,1}M_{\rm BH,2}/M_{\rm BBH}^2$, its frequency $f=nf_{\rm orb}$ with $f_{\rm orb} = (GM_{\rm BBH})^{1/2}/(2\pi a^{3/2})$, and the angle-averaged GW strain means the RMS of the GW strain averaged over the phase angles and the inclination angles. If fixing $\iota=0$ without averaging the inclination angle, the GW amplitude is 
\begin{equation}
H_{n,\iota=0}=8\sqrt{g(n,e)}\frac{G^{5/3}\mathcal{M}^{5/3}}{c^4r n}(2\pi f_{\rm orb})^{2/3},
\label{eq:Hn}
\end{equation}
which is the largest one among those obtained by fixing different $\iota$ values.

In the special case of a circular BBH, only the 2nd harmonic component $H_{2}$ does not vanish ($H_n|_{n\neq 2}=0$), and the GW radiation is almost monochromatic with a frequency being twice the orbital ones, i.e., $f=2 f_{\rm orb}$. Thus, the amplitude of the GW with $\iota=0^{\rm o}$ can be simply expressed as
\begin{equation}
H_{2,\iota=0}=\frac {4{G}^{5/3}\mathcal{M}^{5/3}(\pi f)^{2/3}}{{c}^4r}.
\label{eq:h_0f}
\end{equation}

The timescale for an eccentric BBH with semimajor axis $a$ and eccentricity $e$ decaying to merger by the GW radiation is \citep{1964PhRv..136.1224P,maggiore2008gravitational},
\begin{eqnarray}
\tau_{\rm GW}(a,e) & = & 2.87\,{\rm Myr} \left(\frac{T_{\rm orb}}{\rm yr} \right)^{8/3} \left( \frac{4.26\times 10^6M_\odot}{M_{\rm BBH}}\right)^{5/3} \nonumber \\ 
& & \times \frac{(1+q)^2}{q}  F(e),
\label{eq:taugw}
\end{eqnarray}
where $T_{\rm orb}$ is the orbital period of the BBH,
\begin{equation}
F(e)=\frac{49}{19} \frac{1}{\tilde{a}^4(e)} \int^{e}_0 de' \frac{\tilde{a}^4(e')(1-e'^2)^{5/2}}{e'\left(1+\frac{121}{304}e'^2\right)},
\end{equation}
and
\begin{equation}
\tilde{a}(e)=\frac{e^{12/19}}{1-e^2} \left(1+\frac{121}{304} e^2 \right)^{870/2299}.
\end{equation}
As seen from Equation~\eqref{eq:taugw}, $\tau_{\rm GW} \propto 1/q$ for a given $T_{\rm orb}$ when $q\ll1$, and the GW decay timescale increases with decreasing mass ratios.

\section{Characteristic Strain for circular BBHs}
\label{sec:GWstrain}

The physical quantity to describe the PTA sensitivity is normally the characteristic GW strain, the power spectra, or the energy density. In this paper, we adopt the characteristic strain $h_{\rm c}$ for S/N estimates. In this section, we review the formulism of $h_{\rm c}$ for the basic case of circular BBHs.

The characteristic GW strain of an inspiralling BBH on a circular orbit can be defined as follows if using the matched-filtering method \citep[see][]{2015CQGra..32a5014M, 2005ApJ...623...23S}
\begin{equation}
h_{\rm c}(f)\equiv \sqrt{2N}H_{2,\iota=0},
\label{eq:hccir}
\end{equation}
where $N$ is the number of the cycles near frequency $f$, $H_{2,\iota=0}$ is the GW amplitude in Equation~\eqref{eq:h_0f}. The dependence of the GW wave on viewing angle is considered in the geometrical factor $\chi$ in Section~\ref{sec:SNR}. At the high-frequency band, the number of cycles generated around $f$ is roughly given by $N_{\rm cyc}= f^2 / \dot{f}\propto f^{-5/3}$ and $N_{\rm cyc} T_{\rm orb} <T_{\rm obs}$, where the frequency variation rate $\dot{f}=\frac{96}{5}\pi^{8/3} \left(\frac{G\mathcal{M}}{c^3} \right)^{5/3}f^{11/3}$, $T_{\rm orb}$ is the time duration of a single cycle, $T_{\rm obs}$ is the observation duration, and thus we adopt $N=N_{\rm cyc}$ in the above equation. At the low-frequency band, however, $N_{\rm cyc} T_{\rm orb} > T_{\rm obs}$, we thus adopt $N=T_{\rm obs} f$. Therefore, the characteristic GW strain can be given by 
\begin{equation}
h_{\rm c}(f)=\sqrt{2}H_{2,\iota=0} \min\left\{\sqrt{\frac{f^2}{\dot{f}}}, \sqrt{T_{\rm obs}f}\right\}.
\label{h_cff}
\end{equation}

Combining Equations~\eqref{eq:h_0f} and \eqref{h_cff} we have
\begin{equation}
\label{h_cf}
h_{\rm c}(f)=\left\{
\begin{aligned}
&\frac{\sqrt {15}{G}^{5/6}\mathcal{M}^{5/6}}{3{\pi }^{2/3} {c}^{3/2}r} f^{-1/6}, \enspace &\text{if $N_{\rm cyc}\le T_{\rm obs}f$};\\
&\frac{4\sqrt{2}\pi^{2/3}T^{1/2}_{\rm obs}G^{5/3}\mathcal{M}^{5/3}}{{c}^4r} f^{7/6}, \enspace &\text{if $N_{\rm cyc}>T_{\rm obs}f$}.
\end{aligned}\right.
\end{equation}

For those sources to be detected by PTAs in the frequency range from $\mu$Hz to nHz, we usually have $N_{\rm cyc}T_{\rm orb} > T_{\rm obs}$ (except for those BBHs with total mass $\ga 10^{10}M_{\odot}$), and thus we can adopt the second line of Equation~\eqref{h_cf} to estimate the characteristic strain. If $T_{\rm obs}=10$\,yr, the characteristic GW strain from a hypothetical BBH in the GC is roughly
\begin{equation}
\begin{split}
h_{\rm c} \approx & 1.9\times10^{-14} \frac{q}{(1+q)^{2}} \left(\frac{f_0}{10^{-8}\rm Hz}\right)^{\frac{7}{6}}\\
& \times \left(\frac{T_{\rm obs}}{10\rm yr}\right)^{\frac{1}{2}} \left(\frac{8\rm kpc}{r}\right)\left(\frac{M_{\rm BBH}}{4.26\times10^6M_\odot}\right)^{\frac{5}{3}}.
\end{split}
\label{eq:hc}
\end{equation}

If the BBH is on an eccentric orbit, we will define an effective characteristic strain given in Section~\ref{subsec:h_cp} below by considering eccentric BBH S/N estimates of PTA observations.

\section{PTA Observations}
\label{sec:PTA}

The detectability of BBH sources by PTA depends on the PTA properties, especially the timing noise of PTA MSPs, and also related to the method adopted in the data analysis, which are summarized below.

\subsection{Noise and properties of PTAs}
\label{subsec:PTA}

The noise for PTA to detect individual sources comes mainly from three parts: the shot noise, the red intrinsic spin noise, and the confusion with GW background (GWB) \citep{2015MNRAS.451.2417R, 2019MNRAS.485..248G, 2020ApJ...897...86C}. The red intrinsic spin noise is quite uncertain \citep{2020MNRAS.497.3264G,2016MNRAS.458.2161L, 2023ApJ...951L..10A, 2023A&A...678A..49E}, thus we do not consider it in this paper. The power spectrum density (PSD) of shot noise can be written as \citep{2011gwpa.book.....C}
\begin{equation}
S_{\rm n,s}(f)=8\pi^2\sigma_{\rm t}^2 f^2\Delta t,
\end{equation}
where $\sigma_{\rm t}$ is the RMS residual of the white noise in the time of arrival (TOA) of pulses from MSPs and $\Delta t$ is the cadence of the PTA observations. 

The characteristic strain of the GWB can be written as \citep[e.g.,][]{2020ApJ...897...86C}
\begin{equation}
h_{\rm b}=\mathcal{A}\frac{(f/1{\rm yr}^{-1})^{-2/3}}{[1+(f_{\rm bend}/f)^{\kappa_{\rm  gw}\gamma_{\rm gw}}]^{1/(2\gamma_{\rm gw})}}. 
\end{equation}
According to recent observations by NANOGrav, we adopt $h_{\rm b}=2.4\times 10^{-15}$ at $f=1{\rm yr}^{-1}$  \citep{2023ApJ...955..132C}. The parameters $f_{\rm bend}$, $\kappa_{\rm gw}$, and $\gamma_{\rm gw}$ could not be directly obtained by current observations of the GWB spectrum; for simplicity, we adopt $f_{\rm bend} =1.15\times 10^{-10}$\,Hz, and $\kappa_{\rm gw}=3.70$, $\gamma_{\rm gw}=0.19$ as those modeled in \citet{2020ApJ...897...86C}. According to those settings, we obtain $\mathcal{A}=2.5\times 10^{-15}$. (For the strain of GWB, see also \citealt{2015PhRvD..91h4055S, 2019MNRAS.488..401C}). The total noise for individual sources may be described in the PSD form as
\begin{equation}
S_{\rm n}(f)=S_{\rm n,s}+\frac{h_{\rm b}^2}{f},
\end{equation}
or in the strain form as
\begin{equation}
h_{\rm n}(f)=\sqrt{fS_{\rm n,s}+h_{\rm b}^2}.
\label{eq:h_nB}
\end{equation}
If the GWB can be well detected, then it can be taken as a known signal and removed from the noise, thus 
\begin{equation}
h_{\rm n}(f)=\sqrt{fS_{\rm n,s}}.
\label{eq:h_nnoB}
\end{equation}
The GW frequency range that can be probed by a PTA is limited by the cadence $(\Delta t)$ and the total observation period $(T_{\rm obs})$, i.e., $1/\Delta t \gtrsim f \gtrsim 1/T_{\rm obs}$. Current PTAs normally set a cadence of $\Delta t \sim 1-2$\,weeks and have been running for a total observation period of $T_{\rm obs} \sim 10-20$\,years. Therefore, the frequency range that the current PTAs can probe is roughly $3-100$\,nHz.

The RMS magnitude of the shot noise ($\sigma_{\rm t}$) is important in determining the sensitivity of a PTA. Different PTAs may have different $\sigma_{\rm t}$, for example, $\sigma_{\rm t}\sim 100$\,ns for current PTAs, $\sigma_{\rm t}\sim 20$\,ns for the CPTA with FAST \citep[][]{2016ASPC..502...19L}, and $\sigma_{\rm t}\sim 10-30$\,ns expected by future SKA-PTA \citep[e.g.,][]{2010CQGra..27h4016S}.

In Table~\ref{tab:t1}, we list the assumed settings of several PTAs, including the PTA in the SKA era (SKA-PTA), and hypothetical PTAs, composed of some assumed MSPs in the vicinity of the MBH in the GC, LMC, M31, or M32 (denoted by GC-PTA, LMCC-PTA, M31-PTA, M32C-PTA), where we omit the case of M87 as it would be significantly faraway for pulsar detection. Each current PTA has monitored about $N_{\rm pl} \sim 20-60$ MSPs, while future SKA-PTA is expected to monitor up to $N_{\rm pl}=1000$ MSPs. We also assume that the hypothetical GC-, M31C-, M32C-, and LMCC-PTAs will monitor $N_{\rm pl}=10$, $5$, $5$, and $5$ MSPs in the vicinity of each central MBH, respectively. 

\begin{figure*}
\centering
\includegraphics[width=0.90\textwidth]{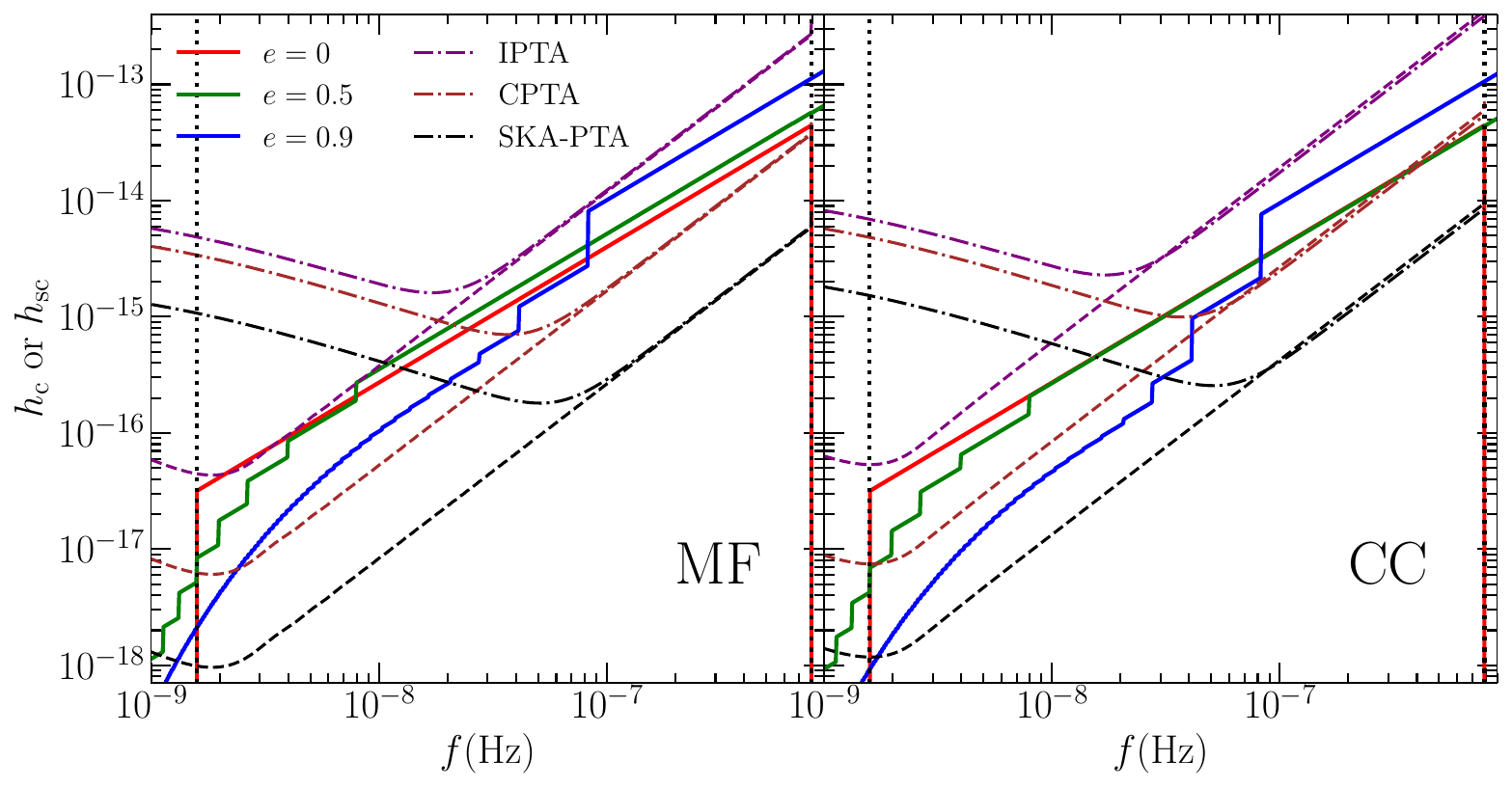}
\caption{
Sensitivity curves ($h_{\rm sc}$) of different PTAs and the effective characteristic strains ($h_{\rm c,eff}$) for GW signals radiated from a hypothetical GC BBH with different eccentricities. Left and right panels represent the results obtained by the matched-filtering (MF) and the cross-correlation (CC) method, respectively. The two vertical black dotted lines in each panel indicate the upper limits and the lower limits for the PTA frequency band. The dashed-dotted lines and the dashed lines represent $h_{\rm sc}$ with and without consideration of the GWB as noise, respectively, where different colors represent different PTAs, and the threshold of S/N  and the total observation time are set as  $\varrho_{\rm th}=1$ and $T_{\rm obs}=20$\,yr in obtaining $h_{\rm sc}$. The dashed-dotted lines are obtained by Equations~\eqref{eq:h_scMF}-\eqref{eq:h_scCC} and the turnover at $\sim3\times10^{-8}$\,Hz is caused by including the GWB in the noise; the dashed lines are obtained by Equations~\eqref{eq:hscMF}-\eqref{eq:hscCC} in Appendix~\ref{sec:eSNR} and have a turnover around the lower limits of the PTA frequency band. These sensitivity curves of $h_{\rm sc}$ shown in this figure are similar to the `frequency domain' curves shown in \citet{2015CQGra..32e5004M}. The solid lines represent $h_{\rm c,eff}$ for GW signals radiated from a hypothetical GC BBH with total mass $4.26\times 10^6M_\odot$ and mass ratio $0.01$, where different colors represent different eccentricities (see Eqs.~\ref{eq:h_cpMF} and \ref{eq:h_cpCC}) and the line with higher eccentricities is higher at the high-frequency end. See details in Section~\ref{sec:PTA}.
}
\label{fig:f2}
\end{figure*}

\begin{table}
\caption{Parameter settings assumed for different PTAs
}
\begin{center}
\begin{tabular}{cccccc}
\hline \hline
PTAs & $N_{\rm pl}$ & $\sigma_{\rm t}$\,(ns) & $T_{\rm obs}$\,(yr) & $\Delta t$\,(yr) & $r_{\rm p-BBH}$\,(pc)\\ \hline
IPTA & 49 & 100 & 20 & 0.04 & $\cdots$\\
CPTA & 100 & 20 & 20  & 0.04 & $\cdots$\\
SKA-PTA & $10^3$ & 10 & 20 & 0.04 & $\cdots$ \\
GC-PTA  & 10 & 100 & 20 & 0.02 & 1 \\
M31C-PTA & 5 & 100 & 10 & 0.02 & 1 \\
M32C-PTA & 5 & 100 & 10 & 0.02 & 1 \\ 
LMCC-PTA & 5 & 100 & 10 & 0.02 & 0.1 \\ \hline \hline
\end{tabular}
\end{center}
\tablecomments{
Columns from left to right denote the notation of the PTAs (see Section~\ref{subsec:PTA}), total number of MSPs monitored by the PTA $N_{\rm pl}$, the timing precision $\sigma_{\rm t}$, the mean cadence $\Delta t$, and the mean distance of the MSPs to the central GW source is small for some special PTAs. Note that the number of MSPs in IPTA adopted here refers to the first data release of IPTA \citep{10.1093/mnras/stw347}. The number of pulsars in the later second data release of IPTA increases to $65$ \citep{2022MNRAS.510.4873A}. 
}
\label{tab:t1}
\end{table}

\subsection{SNR}
\label{sec:SNR}

The S/N of the GW signal from a hypothetical BBH in the GC monitored by a PTA can be estimated once the PTA noise/sensitivity curve is given (for example, see Fig.~\ref{fig:f2}). Below we adopt two different methods to estimate S/N for PTA observations, i.e., the matched-filtering and the cross-correlation methods. One may note that the difference between the S/Ns given by the matched-filtering method and the cross-correlation method depends on the definition of the S/N. The former is defined to be proportional to GW strain amplitude, while the latter is proportional to GW power or squared strain amplitude, which also reflects the difference in the conditions adopted for these two methods \citep[see also][]{maggiore2008gravitational}.
\begin{itemize}

\item {\bf Matched-filtering method:}
if the BBH is on a circular orbit, its GW radiation can be approximated as a monochromatic signal. According to \citet[][Eq.~85 therein]{2022ApJ...939...55G} and \citet{2015CQGra..32a5014M, 2015CQGra..32e5004M}, the S/N $\varrho$ of such a monochromatic GW signal detected by a given PTA can be roughly estimated as
\begin{equation}
\varrho_{\rm MF}=N^{1/2}_{\rm pl}\frac{\chi h_{\rm c} }{h_{\rm n}},
\label{eq:SNR_MFc}
\end{equation}
where $h_{\rm c}$ is from Equations~\eqref{h_cff}, $h_{\rm n}^2=fS_{\rm n}$, and $\chi$ is a geometric factor that is roughly $0.365$ in the far-field approximation (or in the case that the source distance is larger than the PTA pulsar distances by a factor of more than $3-4$) but  deviates from $0.365$ significantly if the GW source distance $r$ is comparable to the distances ($L$) of the pulsars in the PTA \citep[see][]{2022ApJ...939...55G}.
 
If the BBH is on an eccentric orbit, it radiates GWs at multiple discrete frequencies (see Section~\ref{sec:GWsignal}). By adopting the matched-filtering method, the S/N $\varrho$ of this GW signal can be roughly estimated as
\begin{eqnarray}
\varrho_{\rm MF} &\simeq &\left( \sum_{i=n_{\rm min}}^{n_{\rm max}} N_{\rm pl} \frac{2\chi^2 H_{i,\iota=0}^2T_{\rm obs}}{S_{\rm n}(f_i)}\right)^{1/2} \nonumber \\
 & = & N^{1/2}_{\rm pl} \chi \left(\sum_{i=n_{\rm min}}^{n_{\rm max}}\frac{h_{{\rm c},i}^2(f_i)}{h_{\rm n}^2(f_i)}\right)^{1/2}, 
\label{eq:SNR_MFe}
\end{eqnarray}
where the characteristic strain contributed by the $i$th component is $h_{{\rm c},i}(f_i)\equiv H_{i,\iota=0}\sqrt{2f_iT_{\rm obs}}$, and $h_{\rm n}^2(f_i)=f_iS_{\rm n}(f_i)$, $n_{\rm min}=\left[\frac{f_{\rm min}}{f_{\rm orb}}\right]+1$, $n_{\rm max}= \left[\frac{f_{\rm max}}{f_{\rm orb}}\right]$, $f_{\rm min}=1/T_{\rm obs}$, $f_{\rm max}=1/\Delta t$, and $[x]$ represents the maximum integer not larger than $x$. The summation in the above equation is over all the frequency components within the PTA band. 
 
\item {\bf Cross-correlation method:} if adopting this method, the S/N of a monochromatic GW signal from a circular BBH can be estimated as Equation (87) in \citet[][]{2022ApJ...939...55G}
\begin{equation}
\varrho_{\rm CC}=\left(\frac{N_{\rm pl}(N_{\rm pl}-1)}{4}\right)^{1/2}\frac{\chi^2 h_{\rm c}^2}{h_{\rm n}^2}.
\label{eq:SNR_CCc}
\end{equation}
The S/N given by the cross-correlation method is related to that given by the matched-filtering method ($\varrho_{\rm MF}$) as $\varrho_{\rm CC} = \left( \frac{N_{\rm pl}-1}{4 N_{\rm pl}}\right)^{1/2} \chi \varrho^2_{\rm MF}$ and it is $\sim \chi \varrho^2_{\rm MF}/2$ when $N_{\rm pl}\gg 1$.

For the GW signal from an eccentric BBH with multiple harmonics, its S/N can be estimated as
\begin{eqnarray}
\varrho_{\rm CC} &\simeq & \left(\sum_{i=n_{\rm min}}^{n_{\rm max}} \frac{N_{\rm pl}(N_{\rm pl}-1)}{4} \frac{4\chi^4H_{i,\iota=0}^4T^2_{\rm obs}}{S^2_{\rm n}(f_i)}\right)^{1/2} \nonumber \\
& = &  \chi^2\left(\frac{N_{\rm pl}(N_{\rm pl}-1)}{4}\right)^{1/2}  \left( \sum_{i=n_{\rm min}}^{n_{\rm max}}\frac{h_{{\rm c},i}^4(f_i)}{h_{\rm n}^4(f_i)} \right)^{1/2}.\nonumber \\
\label{eq:SNR_CCe}
\end{eqnarray}
\end{itemize}

\subsection{Effective Characteristic Strain for Eccentric Orbits}
\label{subsec:h_cp}

For eccentric orbits, there are many high frequency harmonics rather than a single frequency component in the GW spectrum. In order to compare the GW strain from its source with the sensitivity curves of PTAs (in Section~\ref{subsec:SC}), we define an effective characteristic strain $h_{\rm c,eff}$ at the frequency $f_{\rm pk} = n_{\rm pk} f_{\rm orb}$ with peak power \citep{2003ApJ...598..419W}, where 
\begin{equation}
n_{\rm pk}=\frac{2\left(1+e\right)^{1.1954}}{\left(1-e^{2}\right)^{1.5}}.
\end{equation}
According to Equations~\eqref{eq:SNR_MFe} and \eqref{eq:SNR_CCe}, the effective characteristic strain is defined  as
\begin{equation}
h_{\rm c,eff}(f_{\rm pk}) \equiv h_{\rm n}(f_{\rm pk})\left(\sum_{i=n_{\rm min}}^{n_{\rm max}}\frac{h_{{\rm c},i}^2(f_i)}{h_{\rm n}^2(f_i)}\right)^{1/2}
\label{eq:h_cpMF}
\end{equation}
for the matched-filtering method, and 
\begin{equation}
h_{\rm c,eff}(f_{\rm pk})\equiv h_{\rm n}(f_{\rm pk})\left(\sum_{i=n_{\rm min}}^{n_{\rm max}}\frac{h_{{\rm c},i}^4(f_i)}{h_{\rm n}^4(f_i)}\right)^{1/4}
\label{eq:h_cpCC}
\end{equation}
for the cross-correlation method. We plot some GW strain $h_{\rm c,eff}$ from BBHs in the GC with $q=0.01$ and $e=0$, $0.5$, $0.9$ in Figure~\ref{fig:f2} (red, green, and blue lines, respectively), which are possible to be detected by the CPTA and SKA-PTA. The GWB is not included in the noise in our calculation of $h_{\rm c,eff}$.

\subsection{The Sensitivity of PTAs}
\label{subsec:SC}

The sensitivity of a PTA to individual monochromatic sources is subject to data analysis method, as well as the properties of the PTA. If we adopt the matched-filtering method, according to our S/N formula in Equation~\eqref{eq:SNR_MFc}, the sensitivity curve of a PTA can be approximated as
\begin{equation}
h_{\rm sc,MF}(f)=\frac{\varrho_{\rm th}h_{\rm n}(f)}{\sqrt{N_{\rm pl}}\chi},
\label{eq:h_scMF}
\end{equation}
where $\varrho_{\rm th}$ is the threshold of S/N. For the cases analyzed in this paper, we adopt the near-field regime rather than the far-field approximation to estimate $\chi$ since the distances of those GW sources are not far enough compared with the MSP distances \citep[see details in][]{2022ApJ...939...55G}

If we adopt the cross-correlation method, according to our corresponding S/N formula in Equation~\eqref{eq:SNR_CCc} we have
\begin{eqnarray}
h_{\rm sc,CC}(f) & = & \left(\frac{4\varrho_{\rm th}^2}{N_{\rm pl} (N_{\rm pl}-1)}\right)^{1/4}\frac{h_{\rm n}(f)}{\chi} \\
& = & \left( \frac{4 N_{\rm pl}}{(N_{\rm pl}-1) \varrho^2_{\rm th}}\right)^{1/4} h_{\rm sc,MF}(f).
\label{eq:h_scCC}
\end{eqnarray}

By adopting these formulas and assuming $\varrho_{\rm th} =1$, the sensitivity curves of these PTAs are plotted in Figure~\ref{fig:f2}. The purple, brown, and black dotted-dashed curves with a turnover at $\sim 3\times10^{-8}$\,Hz in Figure~\ref{fig:f2} represent the sensitivity curves of PTAs by considering the confusion from the GWB (based on Eq.~\ref{eq:h_nB}), while the purple, brown, and black dotted lines represent those without considering the GWB confusion (based on Eq.~\ref{eq:h_nnoB}). The overall shapes for $h_{\rm sc,CC}$ and $h_{\rm sc,MF}$ are the same, and $h_{\rm sc,CC} \simeq \sqrt{2} h_{\rm sc,MF}$ by setting $\varrho_{\rm th}=1$ and $N_{\rm pl} \gg 1$ in Equation~\eqref{eq:h_scCC}. If setting  $\varrho_{\rm th} =2$, then $h_{\rm sc,CC} \simeq h_{\rm MF}$, and if setting a substantially higher S/N threshold ($\varrho_{\rm th} \gg 2$), we have $h_{\rm sc,CC} \ll h_{\rm sc,MF}$ .

\begin{figure*}
\centering
\includegraphics[width=0.7\textwidth]{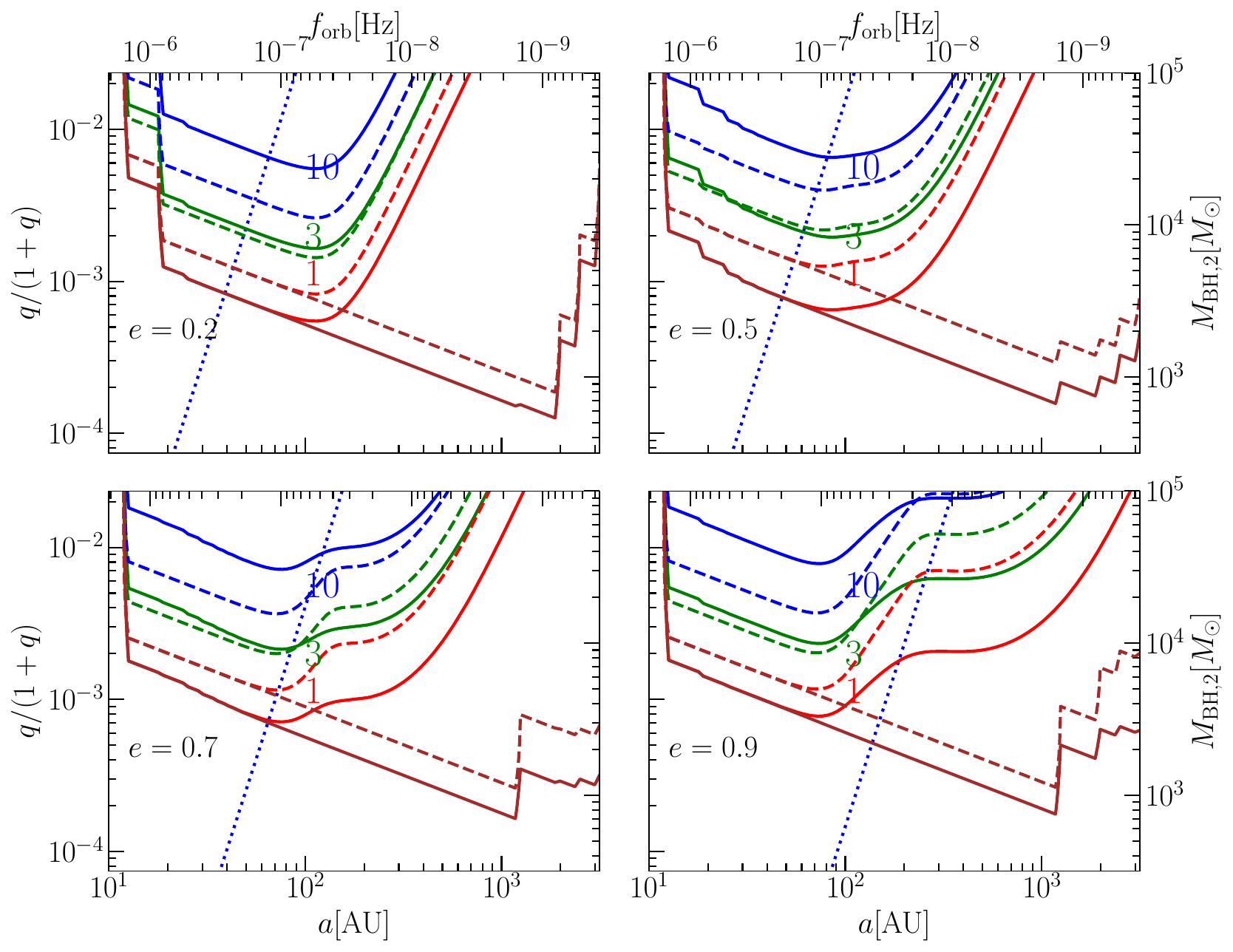}
\caption{Contours of the expected S/N ($\varrho$) in the $q/(1+q)$ versus $a$ parameter space of the hypothetical BBH in the GC, monitored by the assumed future SKA-PTA (see Tab.~\ref{tab:t1}). The hypothetical BBH is assumed to have a mass ratio $q$, semimajor axis $a$, and eccentricity $e$. Top-left, top-right, bottom-left, and bottom-right panels show the cases with $e=0.2$, $0.5$, $0.7$, and $0.9$, respectively. For each panel, the left and the right vertical axes indicate the values of $q/(1+q)$ ($\sim q$ when $q\ll 1$) and the secondary MBH mass $M_{\rm BH,2}=qM_{\rm BBH}/(1+q)$, respectively; the bottom and the top axes indicate the values of $a$ and BBH orbital frequency $f_{\rm orb}$, respectively. Solid and dashed contour lines show the S/N values estimated by adopting the matched-filtering (Eq.~\ref{eq:SNR_MFe}) and cross-correlation methods (Eq.~\ref{eq:SNR_CCe}), respectively. Red, green, and blue colors indicate S/N $\varrho= 1$, $3$, and $10$, respectively, obtained by including the GWB from cosmic BBHs into the noise, while brown color represents S/N $\varrho= 1$, obtained by ignoring the confusion from the GWB. Those BBHs located in the region on the right of the blue dotted lines have the GW merger timescale $\tau_{\rm GW}>10^7$\,yr. This figure suggests that the SKA-PTA can reveal the existence of an IMBH with mass $M_{\rm BH,2} \gtrsim 5\times 10^2- 5\times10^3 M_\odot$ (or $q\gtrsim 10^{-4}-10^{-3}$) and semi-major axis $a \sim 10^2-10^3$\,AU (see brown curves; also see  Fig.~\ref{fig:f5}). The parameter space below the brown curves cannot be ruled out for the existence of an IMBH by this SKA-PTA. See details in Section~\ref{subsec:SKAPTA}.
}
\label{fig:f3}
\end{figure*}

\section{Possible PTA constraints on the existence of a BBH in the GC}
\label{sec:GCMBBH}

\begin{figure*}
\centering
\includegraphics[width=0.7\textwidth]{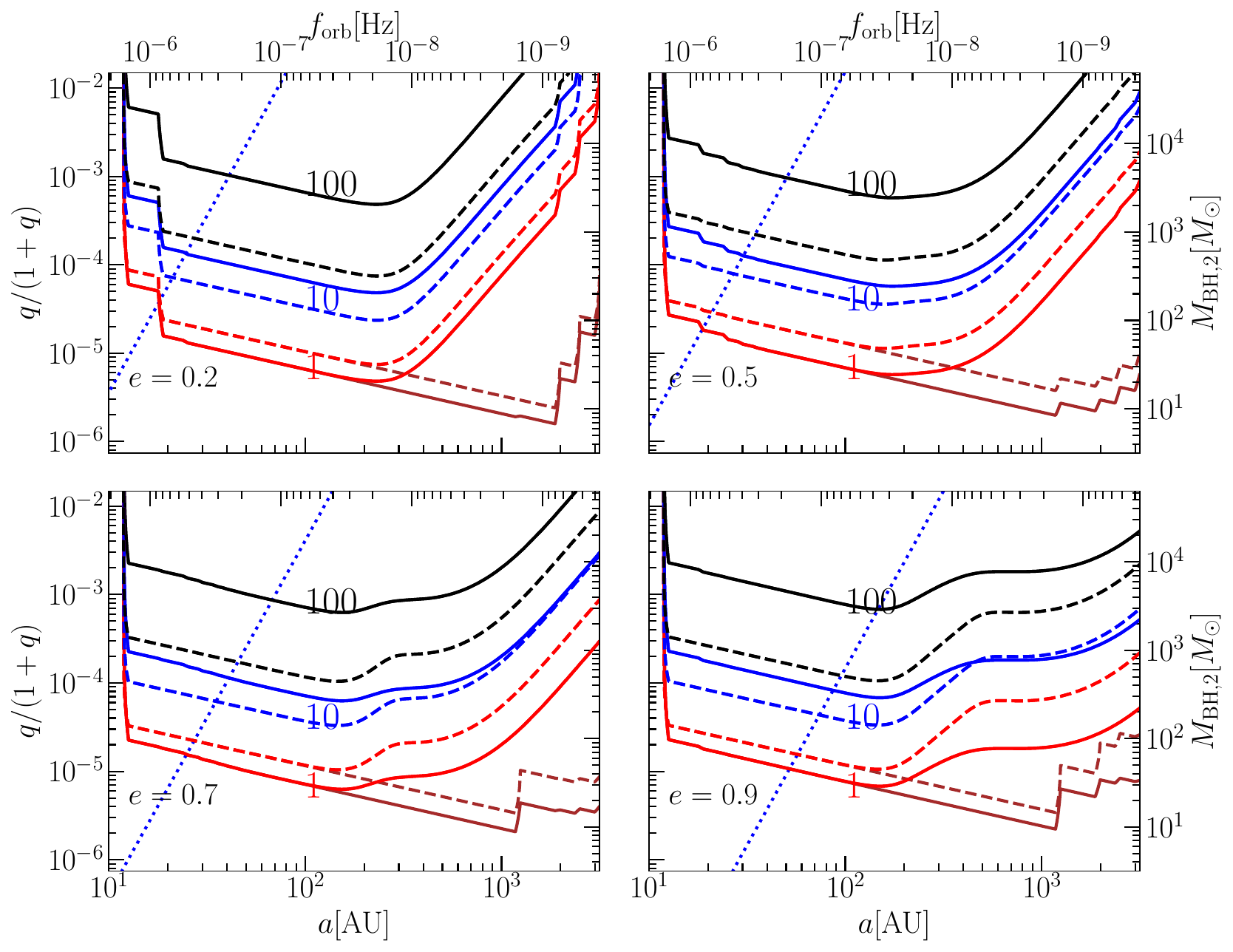}
\caption{
Expected S/N of a hypothetical BBH in the GC with mass ratio $q$, semimajor axis $a$, and eccentricity $e$, monitored by an assumed PTA composed of MSPs in the GC (see GC-PTA in Tab.~\ref{tab:t1}). Legends are similar to those for Figure~\ref{fig:f3}, except that the red, blue, and black lines represent $\varrho= 1$, $10$, and $100$, respectively. This figure suggests that the GC-PTA, if possible, can reveal the existence of a secondary black hole with mass down to a few ten solar masses rotating around the central MBH with a separation of a few ten to several thousand AU. See details in Section~\ref{subsec:GCPTA}.
}
\label{fig:f4}
\end{figure*}

\begin{figure*}
\centering
\includegraphics[width=0.7\textwidth]{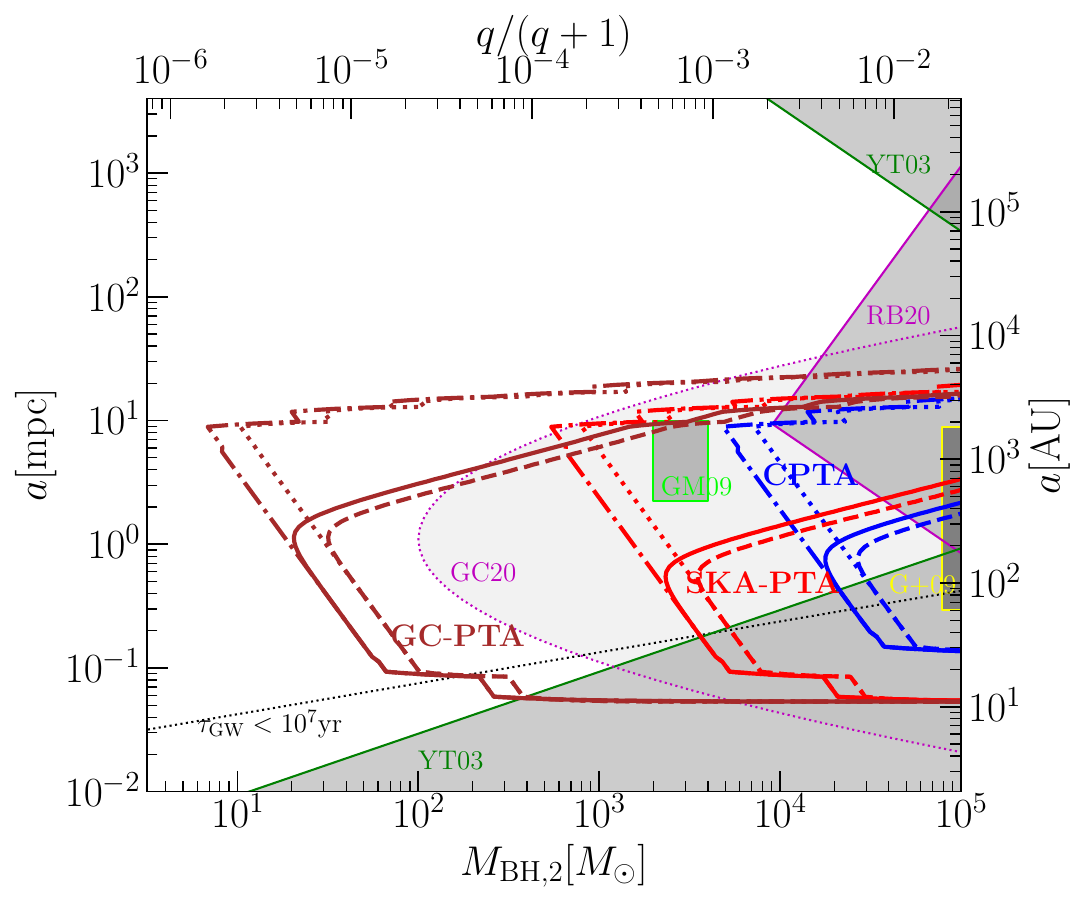}
\caption{
Possible constraints on the parameter space for the existence of a BBH in the GC expected from future PTA observations. The blue, red, and brown lines represent those BBHs ($e=0.2$) with S/N $\varrho=1$ estimated for the CPTA, SKA-PTA, and GC-PTA, respectively, with parameters listed in Table~\ref{tab:t1} by using the MF and the CC methods. The solid and the dash-dotted lines represent the S/N obtained by using the MF method with and without considering the confusion from the GWB, respectively, while the dashed and the dotted lines represent the S/N obtained by using the CC method with and without considering the confusion from the GWB, respectively. In the region with $\varrho>1$ on the right of these cyan/red/brown lines, the BBH could be revealed by the CPTA/SKA-PTA/GC-PTA observations; in the region with $\rho<1$ on the left, the existence of a BBH cannot be ruled out by the corresponding PTA observations. For reference, the limits on the parameter space from other observations and kinematic/dynamical arguments in the literature are also plotted in this figure, and the parameter space in the shaded regions was suggested to be excluded for the existence of a BBH, see details in \citet[][denoted by RB20]{2020ApJ...892...39R} for the region on the right of the magenta solid lines (see also \citealt{2003ApJ...593L..77H}, \citealt{2003ApJ...599.1129Y}, and \citealt{{2004ApJ...616..872R}} for earlier versions), \citet[][YT03]{2003ApJ...599.1129Y} for the region above the dark green solid line at the top-right corner and the region below the dark green solid line at the bottom-right corner, \citet[][GM09]{2009ApJ...705..361G} for the rectangular region bounded by the light green solid lines, \citet[][{G+09}]{2009ApJ...692.1075G} for the region on the right of the yellow solid lines, \citet[][GC20]{2020A&A...636L...5G} for the region on the right of the purple dotted line (for other versions, see also a shrunk irregular region shown in \citealt[][]{2023ApJ...959...58W} and a porous irregular region shown in \citealt[][]{2023A&A...672A..63G}). The black dotted line marks the BBHs with GW decay timescale $\tau_{\rm GW}=10^7$\,yr, below which $\tau_{\rm GW}<10^7$\,yr. See details in Section~\ref{sec:GCMBBH}.
}
\label{fig:f5}
\end{figure*}

In this Section, we investigate the constraint on the parameter space allowed for the existence of a BBH in the GC that may be obtained by future PTA observations. We calculate S/Ns for the hypothetical BBH in the GC with different properties $(q, a, e)$ according to Equations~\eqref{eq:SNR_MFe} and \eqref{eq:SNR_CCe} for different PTAs with the properties assumed in Table~\ref{tab:t1}. We find that the SKA-PTA and GC-PTA can reveal the existence of a hypothetical BBH even if $q\lesssim 0.001-0.0001$ and put a strong constraint on the parameter space of the possible BBH, while the IPTA and CPTA may be able to reveal the existence of the secondary component only if $q \gtrsim 0.02$ (the range of $q$ appears highly unlikely, according to current limits set by some other methods, see \citealt{2003ApJ...599.1129Y, 2003ApJ...593L..77H, 2009ApJ...705..361G, 2009ApJ...692.1075G, 2009ApJ...693L..35M, RevModPhys.82.3121, 2020ApJ...888L...8N, 2020A&A...636L...5G, 2020ApJ...892...39R, 2023MNRAS.525..561E, 2023MNRAS.526.1471F}). Below we list the detailed results obtained for the SKA-PTA and GC-PTA, respectively.

\subsection{SKA-PTA}
\label{subsec:SKAPTA}

Figure~\ref{fig:f3} shows the expected S/N contours ($1$, $3$, and $10$) on the mass ratio $q$ versus semimajor axis $a$ plane (for four different cases of $e$) if the hypothetical BBH in the GC is monitored by the SKA-PTA for $20$ years. For the S/N curves obtained from the matched-filtering method, since $h_{\rm c,eff} \propto q$ when $q\ll 1$, the curves representing S/N $\varrho=3$ and $10$ are almost the same as those by moving the curve $\varrho=1$ upward in the $q/(1+q)$ axis by $0.47$\,dex ($\log 3$) and $1$\,dex. For the curves obtained from the cross-correlation method, as seen from this figure: the parameter space sensitive to the PTA is limited to be $a\sim 10-3\times10^3$\,AU, especially for the low $e$ cases (top-left panel); if $a$ is too large or small, the GW from the BBH is out of the SKA-PTA frequency band and cannot be detected. Compared with the low-$e$ cases, the high-$e$ ones (especially $e=0.9$, bottom-right panel) can have a substantially higher S/N at $a \ga 300$\,AU. The reason is that the high harmonics ($n>2$) with significant power emitted from those high-$e$ BBHs are still in the high-sensitivity frequency range of the PTA. This enables the possibility of revealing a BBH with larger semimajor axis ($a \ga 300$\,AU) and high $e$ by PTA observations. However, the GW decay timescale $\tau_{\rm GW}$ decreases with increasing $e$, therefore, the surviving time for those BBHs with $a\la 100$\,AU is so short ($\ll 10^7$\,yr) that the probability for its existence is substantially lower.  

The GWB may be well modeled and removed from the data analysis as the S/N of the GWB signal increases with observation time, thus the S/N for individual GW sources can be much higher than that resulting from the case with GWB confusion at low frequencies. The brown lines in Figure~\ref{fig:f3} show the expected S/N contours with $\varrho=1$. It is clear that a substantially larger parameter space (especially at $a\sim 20-10^3$\,AU and $M_{\rm BH,2}\gtrsim 5\times 10^2-5\times 10^3M_\odot$) can be constrained with $\varrho>1$. In this case, S/N contours with other values (e.g. $\varrho=10$) are not shown in the figure, but they can be obtained by simply moving the brown lines upward by a fixed factor [e.g., $1$\,dex ($\propto \varrho$) for the MF curve and $0.5$\,dex ($\propto \sqrt{\varrho}$) for the CC curve].

According to the S/N curves shown in Figure~\ref{fig:f3} (brown curves), we conclude that future SKA-PTA observations can put constraints on the parameter space for the existence of an IMBH rotating around the GC MBH as follows. If no GW signal was detected by SKA-PTA within $20$ years of observations, which would suggest independently that there is no IMBH with mass $\gtrsim 500-5\times10^3M_\odot$ (or $q\gtrsim 10^{-4}-10^{-3}$) rotating around the GC MBH with semimajor axis $a\sim 20-3\times10^3$\,AU.

\subsection{GC-PTA}
\label{subsec:GCPTA}

If a number of MSPs located at the GC can be discovered in the future by radio surveys like SKA, these MSPs may also be used to form a PTA (hereafter denoted as GC-PTA) to detect low-frequency GWs, if any, radiated from the GC \citep[see][]{2004ApJ...615..253P, 2012ApJ...752...67K}. This is possible since a number of $\sim 100-1000$ pulsars have been predicted to exist in the vicinity of GC MBH \citep[e.g.,][]{2004ApJ...615..253P, 2014ApJ...784..106Z}. Many efforts have been dedicated to searching for such pulsars in the past decades \citep[e.g.,][]{2010ApJ...715..939M, 2013IAUS..291..382E, 2019ApJ...876...20H}. A magnetar is found at a distance of $\sim 0.1$\,pc from the GC MBH \citep[e.g.,][]{2013Natur.501..391E,2013ApJ...770L..24K}, which indicates that many pulsars should exist in the GC as predicted. Assuming that a GC-PTA formed by $10$ MSPs at a distance $r_{\rm pl}\sim 1$\,pc from Sgr A*, potentially to be found by future radio surveys, we estimate the expected S/N of the GW signals from the hypothetical GC BBH with different semimajor axes, eccentricities, and mass ratios, as done above for the SKA-PTA.

Figure~\ref{fig:f4} shows the expected S/N curves ($1$, $10$, and $100$) on the $q$-$a$ plane (for four different $e$) if the hypothetical GC BBH is monitored by the GC-PTA with a timing RMS noise of $\sigma_t=100$\,ns and a cadence of $\Delta t =0.02$\,yr for $20$ years. Since the assumed GC MSPs are at a distance of only $1$\,pc from the GC BBH, an order of $\sim 10^3-10^4$ times smaller than the distance of Earth (or most MSPs adopted in the SKA-PTA), the GW effect on the GC MSPs should be larger than that on the SKA MSPs by a similar order. For the GC-PTA, the S/N estimate can be obtained by using the same formulas as Equations~\eqref{eq:SNR_MFe} and \eqref{eq:SNR_CCe} except that the geometric factor $\chi$ is much different. Here we have 
\begin{equation}
\chi\approx 0.365\left(\frac{r}{r_{\rm pl}} \right)\simeq 2.92\times10^{3}\left(\frac{r}{8\rm kpc}\right)\left(\frac{r_{\rm pl}}{1\rm pc}\right)^{-1},
\label{eq:chirp}
\end{equation}
if taking $H_i$ in Eqs.~\eqref{eq:SNR_MFe} and \eqref{eq:SNR_CCe} as the GW strain  at Earth with $r=8$\,kpc (Eq.~\ref{eq:Hn}) (for details see Appendix E in \citealt{2022ApJ...939...55G} ). The expected S/N of a given BBH obtained by the GC-PTA is substantially larger than that by the SKA-PTA (see Figs.~\ref{fig:f3} and \ref{fig:f4} for comparison). Therefore, the parameter space of an IMBH rotating around the GC MBH that can be revealed by the GC-PTA is wider compared with that by the SKA-PTA; even an IMBH with mass of a few hundred $M_\odot$, rotating around the GC MBH at $a\sim 10-10^3$\,AU, can be detected by such a GC-PTA with S/N$\sim 10$ (see Fig.~\ref{fig:f4}). 

If the GWB can be well modeled and removed, the S/N for the BBH becomes much larger at larger $a$ and lower orbital frequency $f_{\rm orb}$, as shown by the brown lines in Figure~\ref{fig:f4} for $\varrho=1$, compared with the cases that consider the confusion from the GWB. Therefore, an IMBH with $a\sim 1000$\,AU and $M_{\rm BH,2}$ larger than several tens $M_\odot$ can be revealed.

According to the above S/N estimates, the allowed parameter space for the existence of a BBH in the GC may be well constrained by future PTA observation, especially if a GC-PTA is possible. Figure~\ref{fig:f5} shows the possible parameter space for a BBH (with eccentricity $e=0.2$) in the GC that can be revealed or excluded by the CPTA (cyan lines), SKA-PTA (red lines), and GC-PTA (brown lines), respectively. As seen from this figure, PTAs can mainly put constraint on the existence of a GC BBH with semimajor axis of $a\sim 10-4000$\,AU, which is mostly due to the limitation of the PTA frequency band ($\sim 10^{-7}-10^{-9}$\,Hz, equivalently a period from a week to about ten years). The mass of the MBH companion, if any, can be constrained down to $\sim 5\times 10^3-4\times 10^4M_\odot$ (or $q\sim 10^{-3}-10^{-2}$), $5\times 10^2-5\times 10^3M_\odot$ (or $q\sim 10^{-4}-10^{-3}$), and $\sim 10-10^2M_\odot$ (or $q\sim 2\times 10^{-6}-2\times 10^{-5}$) by the CPTA, SKA-PTA, and GC-PTA, respectively. If the eccentricity of such a BBH is much larger, then the range of the parameter space of an IMBH that can be revealed can be extended to larger $a$ (see Figs.~\ref{fig:f3} and \ref{fig:f4}). It is worthwhile to note that the GW radiating from a $\sim 100M_\odot$ BH (or even smaller objects like S-stars, see \citealt{Cai:2018fkc}) can be detected if a GC-PTA is eventually realized. 

The allowed parameter space for the existence of an IMBH in the GC can also be imited by some dynamical arguments, e.g., on the motion of the GC MBH observed via radio observations, the kinematics of S-stars (including S2), and the stability of S-star-cluster around the GC MBH \citep{2003ApJ...599.1129Y, 2003ApJ...593L..77H, 2020ApJ...892...39R, 2006ApJ...641..319P, 2010MNRAS.409.1146G, RevModPhys.82.3121, 2017ApJ...850L...5T, 2019MNRAS.482.3669G, 2020ApJ...888L...8N, 2020ApJ...892...39R, 2020A&A...636L...5G}. For comparison, in Figure~\ref{fig:f5} we also plot various limits obtained in the previous works based on such dynamical arguments. As seen from this figure, a secondary black hole with a substantially high mass in the GC appears incompatible with the dynamical limits placed by some of those works, and part of the parameter space for the existence of the secondary black hole may be excluded. Compared to those obtained by the dynamical arguments, PTAs, especially the GC-PTA, may give a strong constraint (with high $\varrho$) on the BBH parameter space in the range of $10\,{\rm AU} \lesssim a \lesssim 3000 $\,AU and $100M_\odot \lesssim M_{\rm BH,2} \lesssim 10^4 M_{\odot}$; we emphasize here that PTA detections have the ability to detect IMBHs with even much lower mass, which provide not only independent but also unique probes to the binarity of the MBH in the GC.

\section{Constraining the binarity of MBHs in LMC, M31, M32, and M87}
\label{sec:otherMBBH}

\begin{figure*}
\centering
\includegraphics[width=0.7\textwidth]{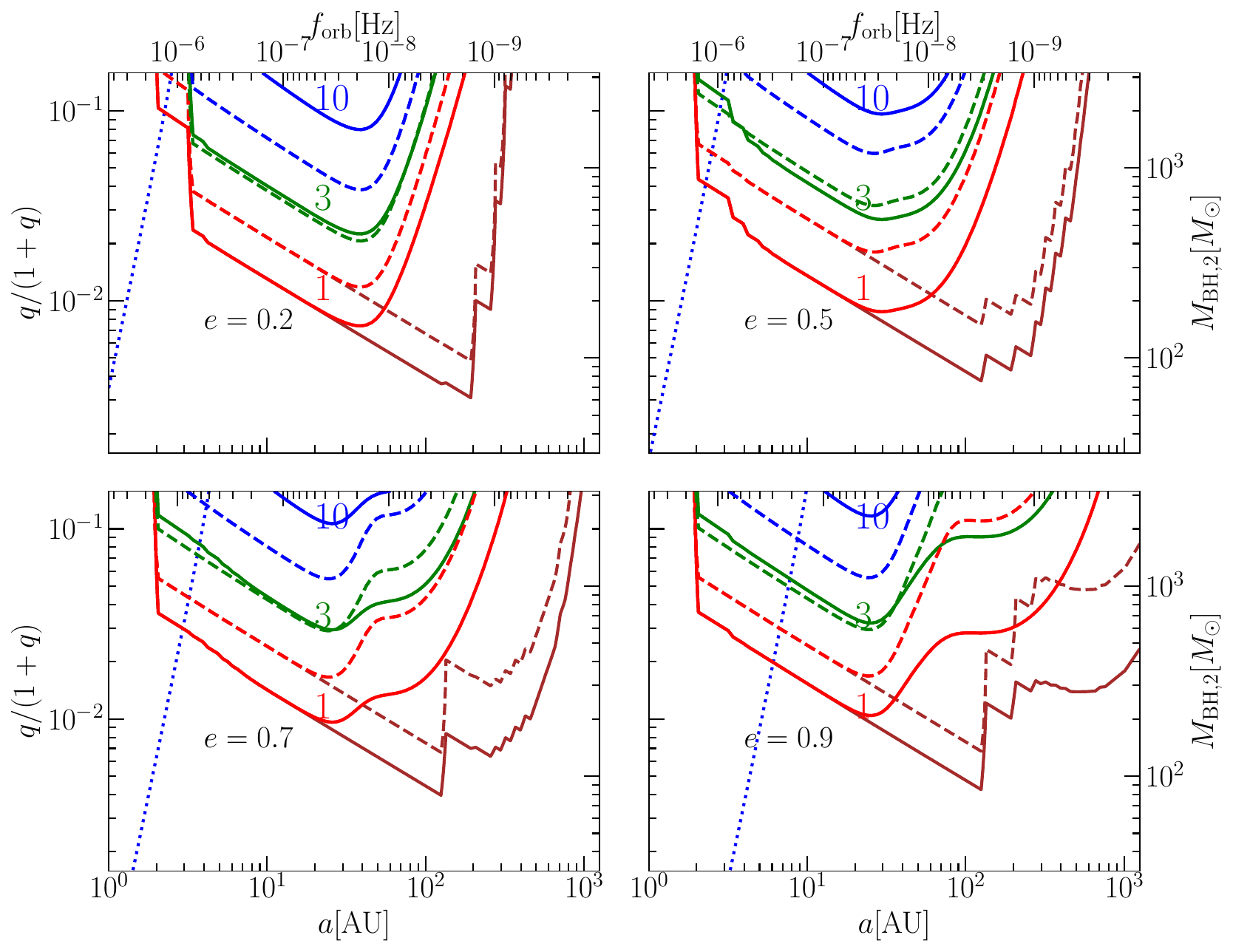}
\caption{Expected S/N of a hypothetical BBH in the LMC center with mass ratio $q$, semimajor axis $a$, and eccentricity $e$ monitored by an assumed PTA composed of MSPs in the LMC center (see LMCC-PTA in Tab.~\ref{tab:t1}). Legends are similar to those for Fig.~\ref{fig:f3}. See details in Section~\ref{subsec:LMC}.
}
\label{fig:f6}
\end{figure*}

It is also possible that a BBH with small mass ratio exists in nearby galactic centers as \citet{2020ApJ...897...86C} predict that a BBH with mass ratio $q \ge  0.01$ and semimajor axis $0<a\le 10$\,pc may survive in the centers of $\gtrsim 10\%$ of nearby galaxies (see Fig.~16 therein). The existence of a BBH in LMC, M31, M32 or M87, is not ruled out by observations, yet, especially if the secondary component is much smaller than the primary one. We investigate the GW radiation from a BBH, if any, in LMC ($r=49.97$\,kpc; \citealt{2013Natur.495...76P}), M31 ($r=774$\,kpc; \citealt{2005ApJ...631..280B}), and/or M32 ($r=805$\,kpc; \citealt{2013ARA&A..51..511K}) and the constraints that can be put by future PTAs on the parameter space for the existence of a secondary MBH rotating around the primary one.

\subsection{LMC}
\label{subsec:LMC}

Observations suggest that an MBH with mass $\sim  2.4\times10^{4} M_\odot$\footnote{Note here the black hole in the LMC is in the mass rage of IMBHs, but we do not distinguish it from an MBH, for simplicity.}  might exist in the center of LMC \citep[e.g.,][]{2017ApJ...846...14B}.  The ejection of a hypervelocity star from LMC also requires at least one MBH existing in the center of LMC \citep[e.g.,][]{2019MNRAS.483.2007E, 2007MNRAS.376L..29G}. Similar to the GC case, we assume that the MBH system in the LMC center is a BBH, with a total mass of $M_{\rm BBH} =2.4\times10^{4} M_\odot$ and an mass ratio $q$. The GW radiation from such a hypothetical BBH in the nano-Hertz frequency range considered in the present paper is much weaker than that from the GC one discussed above because of the smaller black hole mass and the larger distance. According to our calculations of the S/N, the IPTA, CPTA, and SKA-PTA cannot detect a significant signal (with $\varrho>1$) from such a hypothetical BBH in the LMC, thus neither revealing nor excluding and thus the existence of such a BBH. 

If a few MSPs located in the center of LMC can be found by the future SKA survey and adopted to form a LMCC-PTA, we estimate the expected S/N of GW signals from the hypothetical BBH in the LMC center that may be monitored by the assumed LMCC-PTA (see Table~\ref{tab:t1}), where the distance of these MSPs to the hypothetical BBH is assumed to be $r_{\rm pl} \sim 0.1$\,pc. Note that the farthest pulsars detected by current radio telescopes is at a distance of $59.7$\,kpc away from the Earth \citep{2005AJ....129.1993M}\footnote{http://www.atnf.csiro.au/people/pulsar/psrcat}, larger than the LMC distance, and $21$ pulsars in the LMC have already been discovered \citep{2019ARA&A..57..417C}. Figure~\ref{fig:f6} shows the resulting S/Ns on the $q$-$a$ plane for four eccentricities $e=0.2$, $0.5$, $0.7$, and $0.9$, respectively. As seen from this Figure, if $a$ is about a few AU to a few hundred AU and the secondary component has a mass $\gtrsim 10^3M_\odot$, it can be detected with a S/N $\varrho\geq3$. The parameter space for the existence of a BBH in the LMC center can be strongly constrained ($\varrho\ge 1$) to be down to mass $\sim100M_\odot$ (or $q\sim 5\times 10^{-2}$) if $a$ is about a hundred AU.

Note that we do not show the results obtained for IPTA, CPTA, and SKA-PTA here because the expected S/Ns are much smaller than $1$, which neither reveal nor exclude the existence of the hypothetical BBH. 

\subsection{M31}
\label{subsec:M31}

\begin{figure*}
\centering
\includegraphics[scale=0.45]{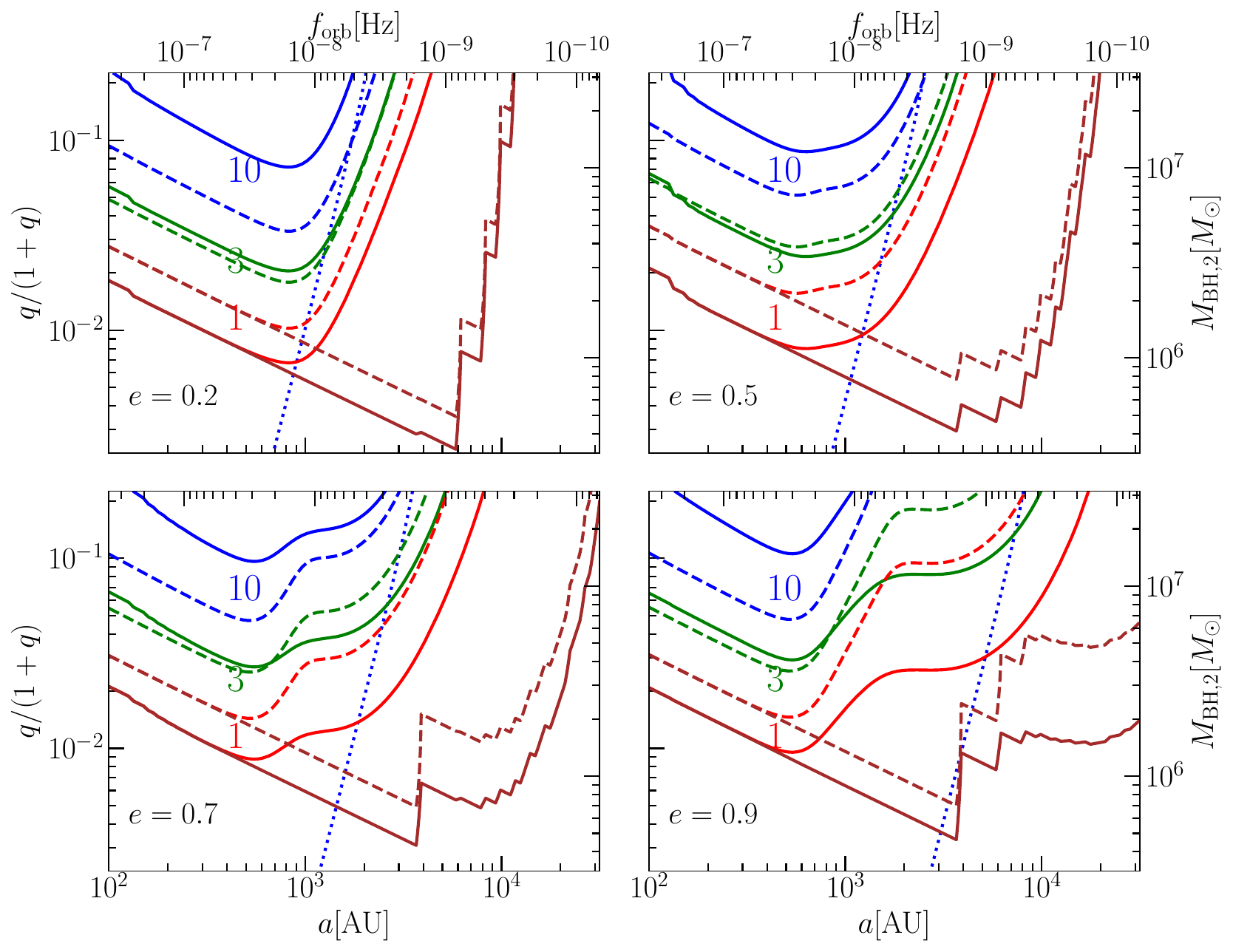}
\caption{Expected S/N of a hypothetical BBH in the center of M31 with mass ratio $q$, semimajor axis $a$, and eccentricity $e$, monitored by the IPTA (see Tab.~\ref{tab:t1}). Legends are similar to those for Fig.~\ref{fig:f3}. See details in Section~\ref{subsec:M31}.
}
\label{fig:f7}
\end{figure*}

\begin{figure*}
\centering
\includegraphics[scale=0.45]{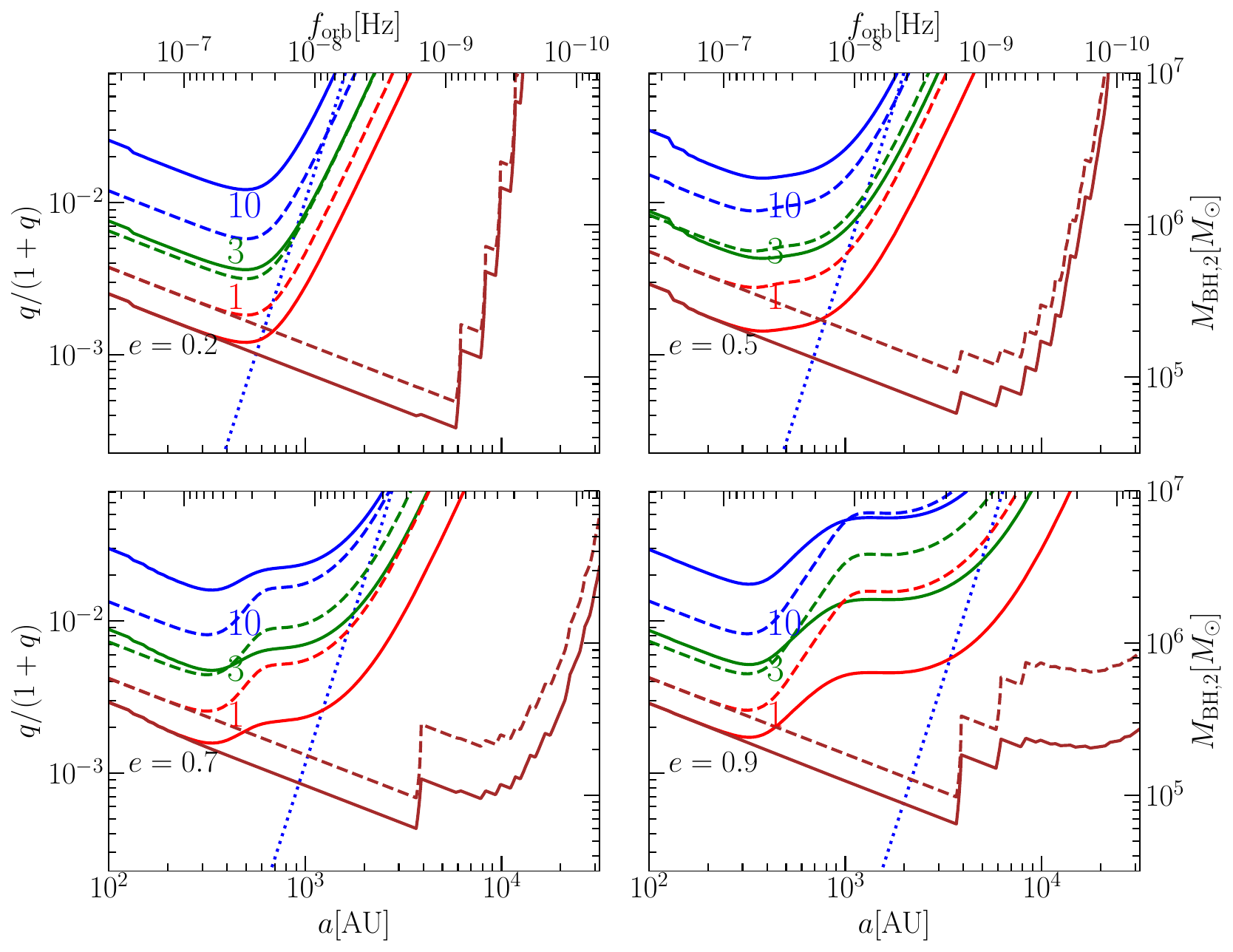}
\caption{Expected S/N of a hypothetical BBH in the center of M31 with mass ratio $q$, semimajor axis $a$, and eccentricity $e$, monitored by the CPTA (see Tab.~\ref{tab:t1}). Legends are similar to those for Fig.~\ref{fig:f3}.  See details in Section~\ref{subsec:M31}.
}
\label{fig:f8}
\end{figure*}

\begin{figure*}
\centering
\includegraphics[scale=0.45]{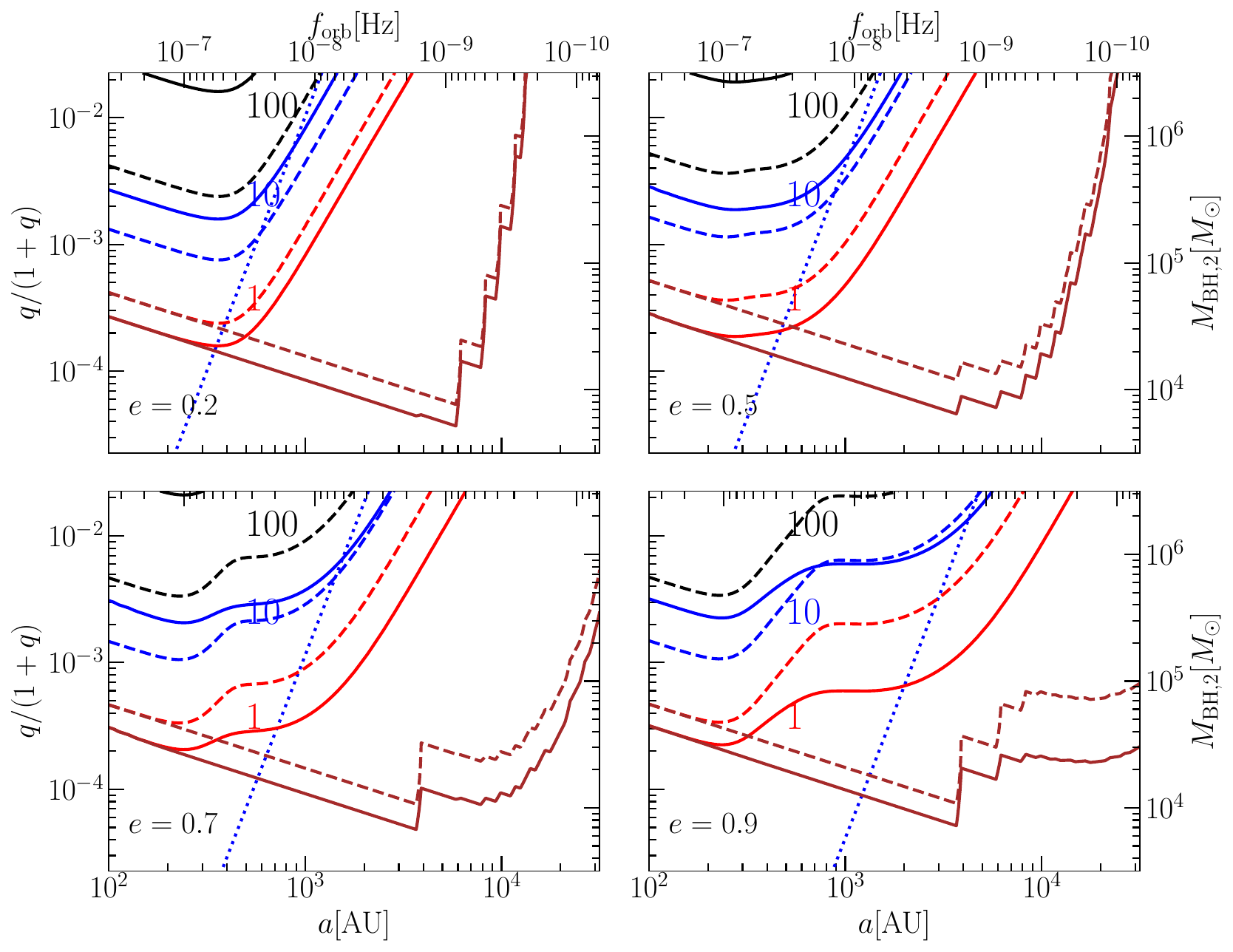}
\caption{Expected S/N of a hypothetical BBH in the center of M31 with mass ratio $q$, semimajor axis $a$, and eccentricity $e$, monitored by the SKA-PTA (see Tab.~\ref{tab:t1}). Legends are similar to those for Fig.~\ref{fig:f3}, except that the red, blue, and black lines represent $\varrho= 1$, $10$, and $100$, respectively. See details in Section~\ref{subsec:M31}.
}
\label{fig:f9}
\end{figure*}

\begin{figure*}
\centering
\includegraphics[scale=0.45]{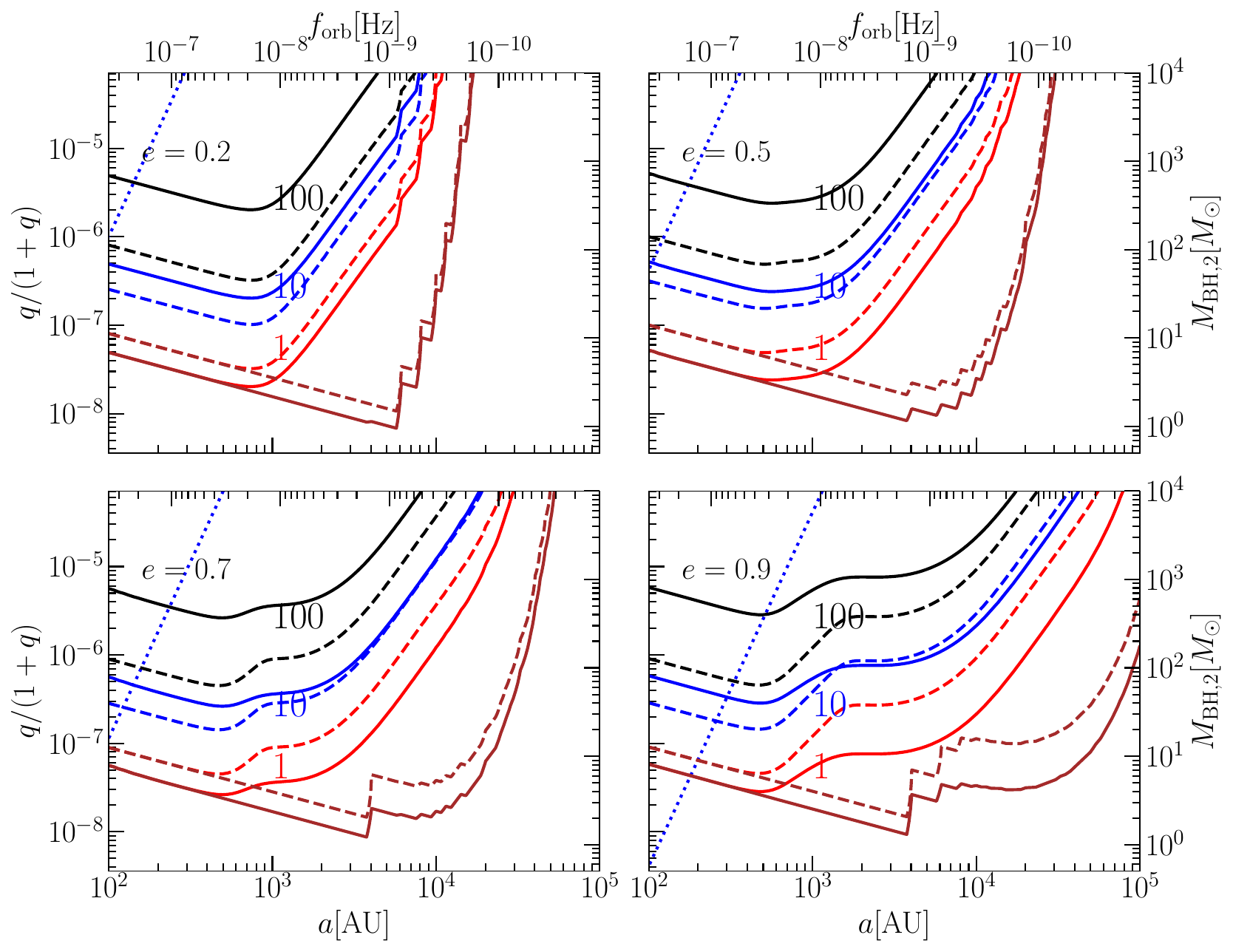}
\caption{
Expected S/N of a hypothetical BBH in the M31 center with mass ratio $q$, semimajor axis $a$, and eccentricity $e$, monitored by an assumed PTA composed of MSPs in the M31 center (see M31C--PTA in Tab.~\ref{tab:t1}). Legends are similar to those for Fig.~\ref{fig:f3}, except that the red, blue, and black lines represent $\varrho= 1$, $10$, and $100$, respectively. See details in Section~\ref{subsec:M31}.
}
\label{fig:f10}
\end{figure*}

Observations have confirmed the existence of at least one MBH with mass $\sim 1.4 \times 10^{8}M_\odot$ in the center of M31 \citep{2005ApJ...631..280B, 2013ARA&A..51..511K}. We hypothesize that a BBH exists in the M31 center, with total mass $M_{\rm BBH}= 1.4 \times 10^{8}M_\odot$ and mass ratio $q$. The GW radiation from such a hypothetical BBH in M31 can be estimated once the BBH semimajor axis and eccentricity are set, and the expected S/N of the GW signal from such a BBH that may be detected by different current/future PTAs (IPTA, CPTA, SKA-PTA, and M31C-PTA; see Tab.~\ref{tab:t1}) as shown in Figures~\ref{fig:f7}-\ref{fig:f10}.

\begin{itemize}
\item IPTA:
Since the mass of the MBH (or the hypothetical BBH) in the center of M31 is much larger than that in the GC, even IPTA can reveal the existence of a BBH in the center of M31 within some parameter space. Figure~\ref{fig:f7} shows the expected S/N distribution for BBHs on the $q$-$a$ plane with four different eccentricities ($e=0.2$, $0.5$, $0.7$, and $0.9$, respectively), monitored by the IPTA, with properties as listed in Table~\ref{tab:t1}. As seen from Figure~\ref{fig:f7} (red and brown curves), a BBH with $q\gtrsim 10^{-2}$ and $a \sim 10^2-10^4$\,AU monitored by the IPTA have an S/N of $\gtrsim 1$.

\item CPTA: Figure~\ref{fig:f8} shows the expected S/N contours  for the hypothetical BBH in the $q$-$a$ plane monitored by the CPTA (with properties as listed in Tab.~\ref{tab:t1}). As seen from this figure, a BBH with $q\gtrsim 10^{-2}$ and $a \sim 10^2-10^3$\,AU can be detected by the CPTA with a S/N of $\gtrsim 10$. If no detection, the parameter space for the existence of a secondary MBH rotating around the primary one can be constrained to be $q\lesssim 10^{-2}$ when $a\sim 10^2-10^4$\,AU. However, one should note here that the merger timescale due to the GW radiation of a BBH with such a semimajor axis in the M31 center is quite short (see the dotted line for $\tau_{\rm GW}=10^7$\,yr in each panel of Fig.~\ref{fig:f8}), which means that the probability for the existence of such a BBH is low.

\item SKA-PTA: Figure~\ref{fig:f9} shows the expected S/N distribution for BBHs expected from the SKA-PTA observations (with properties listed in Tab.~\ref{tab:t1}) on the $q$-$a$ plane with four different eccentricities ($e=0.2$, $0.5$, $0.7$, and $0.9$, respectively). As seen from this figure, a BBH with $q\gtrsim 10^{-3}$ (or $\gtrsim 10^{-4}-10^{-3}$) and $a \sim 100-1000$\,AU (or $a\sim 10^2-10^4$\,AU) can be detected by the SKA-PTA with a S/N of $\varrho\gtrsim 10$ (or $\varrho\gtrsim1$). However, one should note here that the merger timescale due to GW radiation of a BBH with such a semimajor axis in the M31 center is quite short (see the dotted line in each panel of Fig.~\ref{fig:f9}), which means that the probability for the existence of such a BBH is low.

\item M31C-PTA:
If there are some MSPs existing in the center of M31 and they are observable by future radio surveys to form a M31C-PTA (see Tab.~\ref{tab:t1}), we estimate the expected S/N of the hypothetical BBH in M31 monitored by this PTA. Figure~\ref{fig:f10} shows the S/N distribution in the $q$-$a$ plane obtained by using the M31C-PTA. As seen from this figure, the estimated S/N can be quite high for a large parameter space. For example, even if the mass of the secondary component is as small as $\sim 100 M_\odot$ when $a$ is smaller than a few thousands AU, the S/N can still be as high as $\gtrsim 10$ (blue curves in Fig.~\ref{fig:f10}), and thus the secondary component can be revealed even if its mass is only $\sim 100 M_\odot$, close to stellar masses, and $a\sim 100-4\times 10^3$\,AU.
\end{itemize}

\subsection{M32}
\label{subsec:M32}

\begin{figure}
\centering
\includegraphics[width=0.45\textwidth]{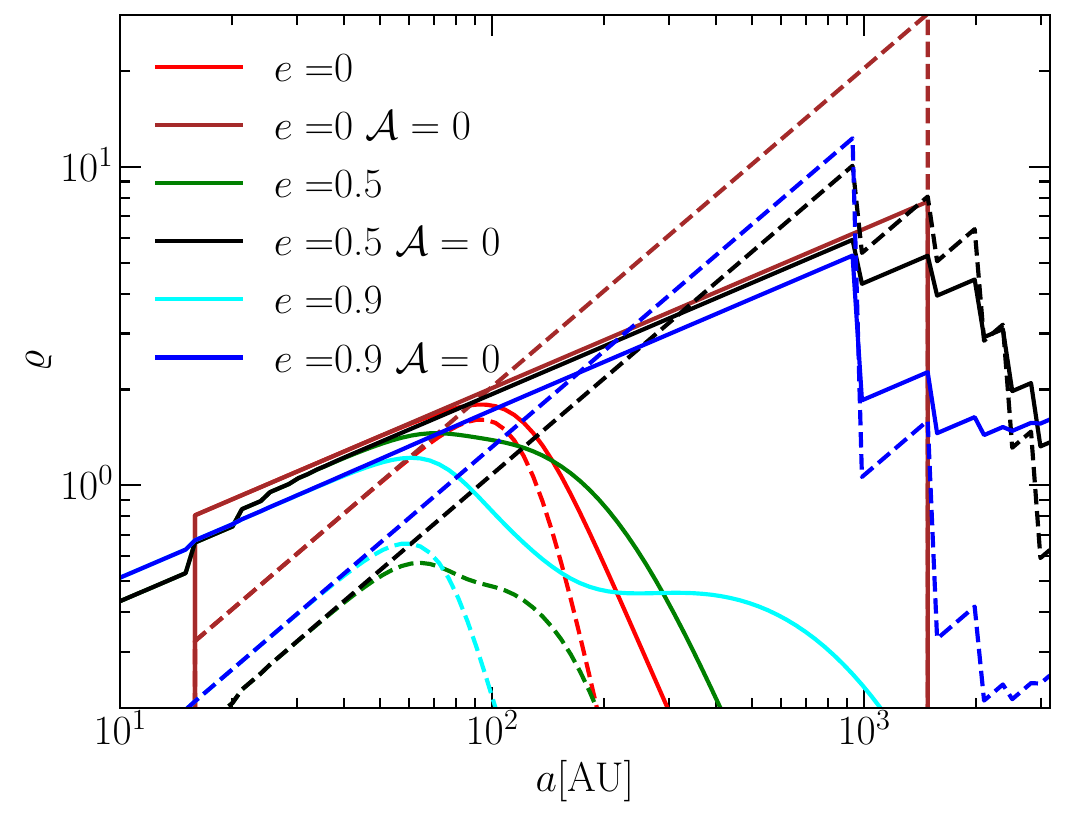}
\caption{Expected S/N of a hypothetical equal-mass BBH in the center of M32 as a function of the orbital semimajor axis $a$, monitored by the SKA-PTA (see Tab.~\ref{tab:t1}). Red, green, and cyan solid (or dashed) lines show the results obtained for a BBH with eccentricity $e=0$, $0.5$, and $0.9$, respectively, by adopting the matched-filtering (MF) (or cross-correlation (CC)) method with consideration of the confusion from the GWB. Brown, black, and blue solid (or dashed) lines show the results obtained for the BBH  with eccentricity $e=0$, $0.5$, and $0.9$, respectively, by adopting the MF (or CC) method without consideration of the confusion from the GWB ($\mathcal{A}=0$). See details in Section~\ref{subsec:M32}.
}
\label{fig:f11}
\end{figure}

\begin{figure*}
\centering
\includegraphics[scale=0.458]{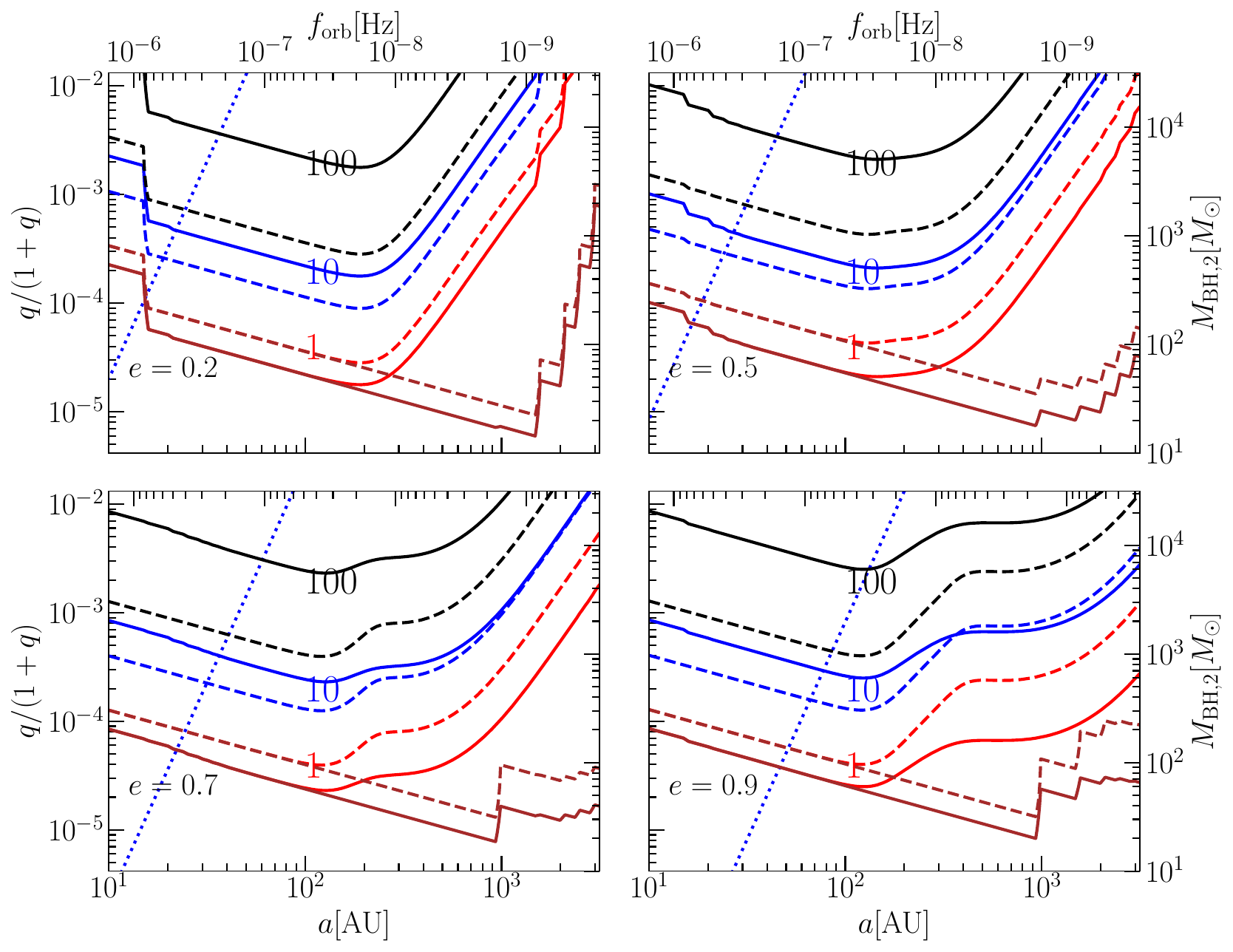}
\caption{Expected S/N of a hypothetical BBH with mass ratio $q$, and semimajor axis $a$ in the center of M32, monitored by a PTA with MSPs in the center of M32 (see M32C-PTA in Tab.~\ref{tab:t1}). The legend is similar to that for Fig.~\ref{fig:f3}, except that the red, blue, and black lines represent $\varrho= 1$, $10$, and $100$, respectively.  See details in Section~\ref{subsec:M32}.
}
\label{fig:f12}
\end{figure*}

Observations have shown that an MBH with mass $2.4\times10^{6} M_\odot$ exists in the center of M32 \citep{2013ARA&A..51..511K}. Similar to the M31 case, we also assume the MBH system in the center of M32 is a BBH with total mass  $M_{\rm BBH} = 2.4\times10^{6} M_\odot$ and mass ratio $q$. We estimate the S/N of the GW signal from such a hypothetical BBH in the M32 center that may be monitored by future SKA-PTA (see Table~\ref{tab:t1}). Figure~\ref{fig:f11} shows the expected S/N as a function of $a$ for equal-mass BBHs with different eccentricities $e=0$, $0.5$, and $0.9$, respectively. As seen from this figure, only when $a\sim 100$\,AU and $e \lesssim 0.5$, can the BBH be detected by future SKA-PTA with an S/N of $\sim 1-2$ if considering the GWB as a noise. For a non-equal-mass BBH with significantly small $q$, the expected S/N would be substantially smaller than $1$, and thus such a low-mass-ratio BBH, if any, in the M32 center is unlikely to be detected by future SKA-PTA. 
If the GWB signal can be well modelled and removed, non-equal-mass BBH with semimajor axis $a\sim10^2-3\times10^3$\,AU and mass ratio $q>0.1$ may have a S/N $\varrho>1$ (see Fig.~\ref{fig:f11}, Eqs.~\ref{eq:SNR_MFe} and \ref{eq:SNR_CCe}, where $h_{\rm c}\propto \mathcal{M}_{\rm c}^{5/3} \propto q$ when $q$ is substantially less than $1$), and thus may be revealed by SKA-PTA.

As assumed for the case of M31 in section~\ref{subsec:M31}, some MSPs are also assumed to exist in the center of M32 and detected by future radio surveys to form a M32C-PTA (see Tab.~\ref{tab:t1} for its settings), we estimate the expected S/Ns of the hypothetical BBHs. Figure~\ref{fig:f12} shows the S/N distribution in the $q$-$a$ plane obtained by using the M32C-PTA. As seen from this Figure, the expected S/N can be $\varrho \gtrsim 10$ if $a\sim 10-10^3$\,AU and $q \gtrsim 10^{-3}$, and $\varrho \gtrsim 1$ if $a\sim 10-10^3$\,AU and $q \gtrsim 10^{-4}$, which suggests that the secondary component with an mass down to $\sim 100M_\odot$ can be revealed at the region with $a\sim 10-10^3$\,AU.

\subsection{M87}
\label{subsec:M87}

Observations have confirmed the existence of at least one MBH with mass $6.5\times10^9M_\odot$ in the center of M87 \citep{2011ApJ...729..119G, 2019ApJ...875L...1E}. We hypothesize that a BBH exists in M87 center, with a total mass $M_{\rm BBH}=6.5\times10^9M_\odot$ and an mass ratio $q$. The GW radiation from such a hypothetical BBH in M87 can be estimated once the BBH semimajor axis and eccentricity are set and the expected S/N of the GW signal from such a BBH that may be detected by future SKA-PTA (see Table~\ref{tab:t1}) can also be estimated. Although the distance of M87 from us  $r\approx16.4$\,Mpc is a little farther than those of local group galaxies \citep{2001ApJ...546..681T}, the MBH mass is large, thus it is also possible to constrain the parameter space of the BBH in M87 using PTA \citep{2016MNRAS.459.1737S,2019MNRAS.488L..90S}. 

\begin{figure*}
\centering
\includegraphics[width=0.7\textwidth]{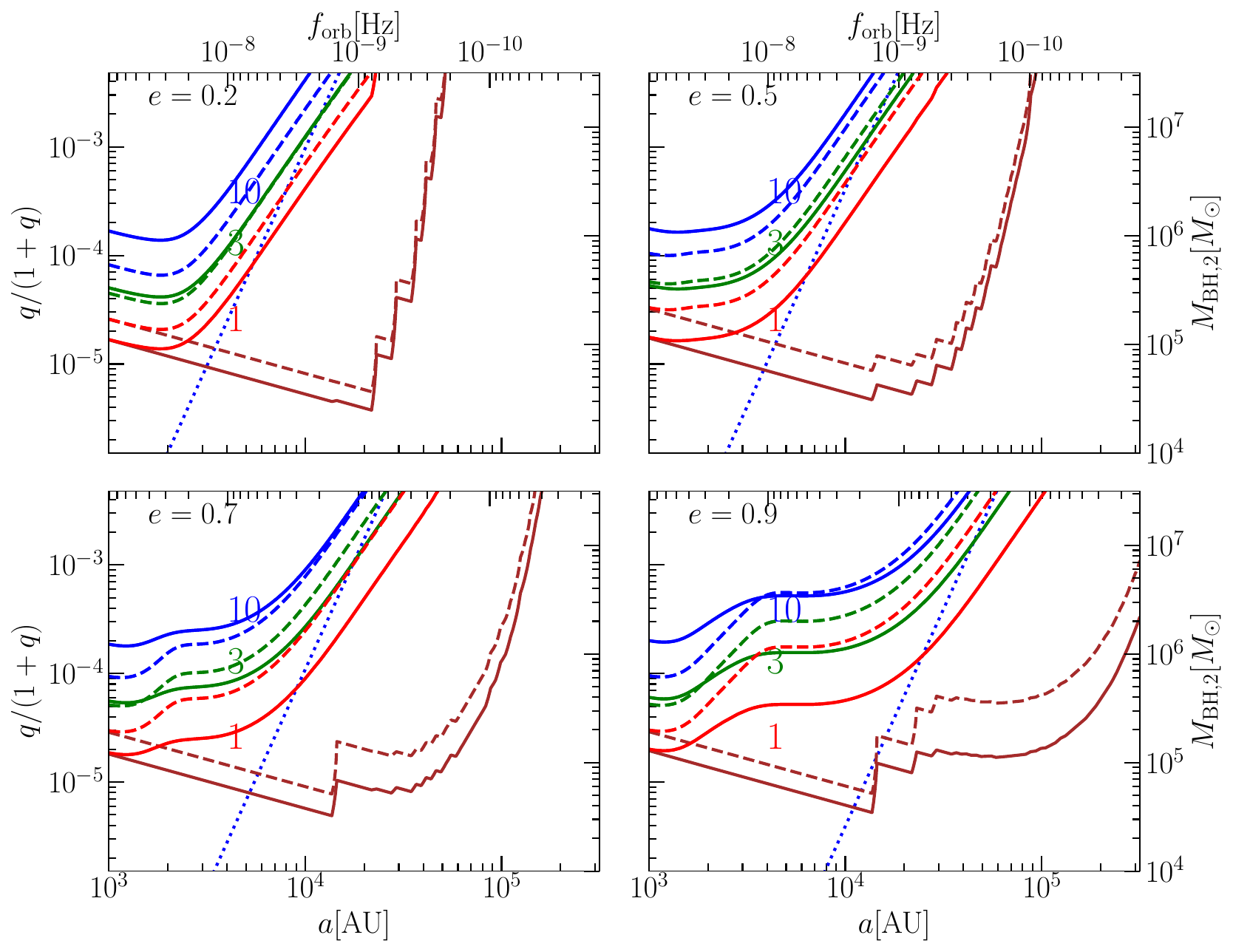}
\caption{Expected S/N of a hypothetical BBH in the center of M87 with mass ratio $q$, semimajor axis $a$, and eccentricity $e$, monitored by the SKA-PTA. The legend is similar to that for Fig.~\ref{fig:f3}. See details in Section~\ref{subsec:M87}.
}
\label{fig:f13}
\end{figure*}

Figure~\ref{fig:f13} shows the expected S/N distribution for BBHs expected from the SKA-PTA observations, with properties listed in Table~\ref{tab:t1}, on the $q$-$a$ plane with four different eccentricities ($e=0.2$, $0.5$, $0.7$, and $0.9$, respectively). As seen from this figure, the expected S/N can be $\varrho \gtrsim 10$ if $a$ is smaller than a few times $10^3$\,AU and $q \gtrsim 10^{-3}$, and $\varrho \gtrsim 1$ if $a\sim 10^3-2\times10^4$\,AU and $q \gtrsim 10^{-5}$. Therefore, the parameter space for the existence of the black hole companion to the MBH in M87 can be well constrained if its semimajor axis is in the range of $a\sim 10^3-2\times10^4$\,AU. One may also note here that the merger timescale due to the GW radiation of BBHs in the M87 center is quite short if $a$ is $\lesssim10^4$\,AU and $q\gtrsim10^{-4}$ (see the dotted line in each panel of Fig.~\ref{fig:f13}), which suggests that the probability for the existence of a BBH with such a semimajor axis is low.

\section{Conclusions}
\label{sec:conclusion}

In this paper, we have investigated whether current and future PTAs can detect the BBHs on circular or eccentric orbits, if any, in the GC and the centers of several nearby typical galaxies (such as LMC, M31, M32, and M87) and the constraints that could be obtained on the binarity of the central MBHs in these galaxies. We have done these by deriving the form for estimating the S/N of eccentric BBHs monitored by PTAs with consideration of the multiple harmonics of the GW spectra, and define an effective characteristic strain for an arbitrary eccentric BBH at the frequency of its peak harmonics, in the cases either using the matched-filtering method or the cross-correlation method to estimate their S/Ns. The above definition of the effective characteristic strain enables an feasible way to demonstrate detectable sources with various orbital parameters in the plots with PTA sensitivity curves. Our main results obtained by applying that formulism are summarized as follows.

\begin{itemize}
\item {\bf GC:} Assuming that there exists an IMBH rotating around the MBH in the GC on a nearly circular orbit, the SKA-PTA or CPTA can detect it with S/N $>1$ within an observation period of $\sim 20$\,year if its mass is about $\gtrsim 3\times 10^3M_\odot$ or $3\times 10^4M_\odot$ and its semimajor axis is $\sim 20-10^3$\,AU. Therefore, the parameter space for the existence of such an IMBH can be independently constrained by future SKA-PTA and CPTA observations. For an IMBH, even with a relatively smaller mass, on an orbit with the same semimajor axis but a larger eccentricity, it can still be detected by the SKA-PTA (or CPTA) as the GW radiation is stronger from a highly eccentric BBH with the same total mass and semimajor axis. 

If a few to ten MSPs could be found in the GC by SKA or some other future radio telescopes and used to form a GC-PTA, then this GC-PTA can detect the GW radiating from an IMBH, if any, with mass from $\sim 10^4M_\odot$ down to $\sim 100M_\odot$ and semimajor axis $\sim 20-10^3$\,AU, rotating around the GC MBH. Therefore, a GC-PTA, if possible, is more powerful in revealing or excluding the existence of IMBH, and the parameter space that cannot be excluded for the BBH existence becomes tighter compared with that constrained by the SKA-PTA and the CPTA.

By comparing the expected constraints on the parameter space for the existence of an IMBH rotating around the GC MBH via PTA observations with the current limits obtained from other methods, we find that the current and future PTA observations can provide a unique way to reveal the IMBH or strongly constrain the allowed parameter space of the IMBH.

\item {\bf M31:} A secondary black hole rotating around the MBH in M31 may be revealed by IPTA, CPTA, and SKA-PTA if the binary has a semimajor axis of $\sim 10^2-10^3$\,AU and a mass ratio $\gtrsim 10^{-2}$, $\gtrsim 10^{-3}$, and $\gtrsim 2\times 10^{-4}$, respectively. If a PTA can be formed by $5$ or more MSPs in the vicinity of the primary MBH, the mass of the secondary that can be revealed may be down to $100M_\odot$ in the semimajor axis range of $\sim 10^2-4\times10^3$\,AU.

\item {\bf M87:} A secondary black hole, if any, rotating around the MBH in M87 may be revealed by the SKA-PTA if the binary has a semimajor axis of $\sim10^3-2\times10^4$\,AU and a mass ratio larger than $\sim 10^{-5}$. 

\item {\bf LMC and M32:} It is hard to detect the secondary BH with $q\ll1$, if any, rotating around the MBH in these two galaxies by using the IPTA, CPTA, or SKA-PTA with a limited observation time (e.g., less than $20$ years), even if the mass ratio of the binary is close to $1$. However, if about $5$ or more MSPs in the vicinity of the MBH in the LMC or M32 can be detected in future and applied as a PTA to probe such a binary, a BBH with mass ratio even down to $\sim 10^{-2}$ or $\sim 10^{-4}$ (and the secondary component with mass down to $\sim 100M_\odot$, close to stellar masses) may be revealed.
\end{itemize}

\acknowledgements{
This work is partly supported by the National Natural Science Foundation of China (grant Nos. 12173001, 12273050, 11991052), the National SKA Program of China (grant no. 2020SKA0120101), the National Key Research and Development Program of China (grant nos. 2022YFC2205201, 2020YFC2201400), the Strategic Priority Program of the Chinese Academy of Sciences (grant no. XDB0550300), and a fellowship of China National Postdoctoral Program for Innovative Talents (grant no. BX20230104). 
}


\appendix 

\section{S/N Formulas}
\label{sec:eSNR}

If there is no confusion from the GWB in the noise, it is straightforward to express the exact S/N formulas as follows, without approximating $\delta_T(f) = \delta(f)$ (here $\delta(f)$ is the Dirac $\delta$ function). For the matched-filtering Method, the exact S/N formulas should be
\begin{equation}
\varrho^2=N_{\rm pl} \frac{\chi^2h_c^2f_0^2}{h_{\rm n}^2T_{\rm obs}}\int^{1/\Delta t}_{1/T_{\rm obs}} df\frac{\delta^2_T(f-f_0)}{f^2},
\end{equation}
where $\delta_T(f)=\frac{\sin(\pi T_{\rm obs}f)}{\pi f}$. We define $I_2(f_0)=\int^{1/\Delta t}_{1/T_{\rm obs}} df\frac{\delta^2_T(f-f_0)}{f^2}$, thus we have
\begin{equation}
\varrho^2=N_p\frac{\chi^2h_c^2f_0^2}{h_{\rm n}^2T_{\rm obs}}I_2(f_0).
\end{equation}
Therefore, we can obtain the sensitivity curve
\begin{equation}
h_{\rm sc}(f)=\left(\frac{T_{\rm obs}}{N_pf^2I_2(f)}\right)^{\frac{1}{2}}\frac{\rho_{\rm th}h_{\rm n}(f)}{\chi}.
\label{eq:hscMF}
\end{equation}

Similarly, the exact S/N formula  for the Cross-Correlation Method is 
\begin{equation}
\varrho^2=\frac{N_{\rm pl}(N_{\rm pl}-1)}{2}\frac{\chi^4h_c^4f_0^4}{2h_{\rm n}^4T^3_{\rm obs}}\int^{1/\Delta t}_{1/T_{\rm obs}} df\frac{\delta^4_T(f-f_0)}{f^4}.
\end{equation}
We also define $I_4(f_0)=\int^{1/\Delta t}_{1/T_{\rm obs}} df\frac{\delta^4_T(f-f_0)}{f^4}$, hence we have
\begin{equation}
\varrho^2=\frac{N_{\rm pl}(N_{\rm pl}-1)}{2}\frac{\chi^4h_c^4f_0^4}{2h_{\rm n}^4T^3_{\rm obs}}I_4(f_0).
\end{equation}
Finally, the sensitivity curve can be solved out as
\begin{equation}
h_{\rm sc}(f)=\left(\frac{4\rho_{\rm th}^2T^3_{\rm obs}}{N_p(N_p-1)f^4I_4(f)}\right)^{\frac{1}{4}}\frac{h_{\rm n}(f)}{\chi}.
\label{eq:hscCC}
\end{equation}

\bibliographystyle{yahapj}
\bibliography{refer}
\end{CJK*}
\end{document}